\documentclass[prd,twocolumn,eqsecnum,showpacs,letterpaper,superscriptaddress,altaffilletter]{revtex4}
\usepackage{amsmath}
\usepackage{color}
\usepackage{graphicx}
\usepackage{epsfig}
\usepackage{latexsym}

\def\be{\begin{equation}}
\def\ee{\end{equation}}
\def\ben{$$}
\def\een{$$}
\def\bea{\begin{eqnarray}}
\def\eea{\end{eqnarray}}
\def\bean{\begin{eqnarray*}}
\def\eean{\end{eqnarray*}}

\def\bi{\begin{itemize}}
\def\ei{\end{itemize}}
\def\ben{\begin{enumerate}}
\def\een{\end{enumerate}}

\def\Ep{{E_+}}
\def\Ip{{I_+}}
\def\Ec{{E_\times}}
\def\Ic{{I_\times}}
\def\Esl{{E_\mathrm{SL}}}
\def\Isl{{I_\mathrm{SL}}}
\def\En{{E_\mathrm{null}}}
\def\In{{I_\mathrm{null}}}
\def\Etot{{E_\mathrm{tot}}}
\def\Esc{{E_\mathrm{soft}}}
\def\hp{{h_+}}
\def\hc{{h_\times}}
\def\xp{{\textsc{X-Pipeline}}}
\def\matlab{{\textsc{matlab}}}

\def\data{\boldsymbol{d}}
\def\fdata{\boldsymbol{\tilde{d}}}

\def\fnoise{\boldsymbol{\tilde{n}}}
\def\h{\boldsymbol{h}}
\def\fh{\boldsymbol{\tilde{h}}}
\def\fhmax{\boldsymbol{\tilde{h}_\mathrm{max}}}

\def\F{\boldsymbol{F}}

\def\Fplus{\boldsymbol{F^+}}
\def\Fcross{\boldsymbol{F^\times}}
\def\FplusDP{\boldsymbol{f}{}^+}
\def\FcrossDP{\boldsymbol{f}{}^\times}
\def\FplusHat{\boldsymbol{e}{}^+}
\def\FcrossHat{\boldsymbol{e}{}^\times}

\def\ct{\dagger}  

\newcommand{\psdk}[1]{S_{#1}[k]}

\def\e{\mathrm{e}}
\def\runone{S5-VSR1}
\def\runtwo{S6-VSR2}

\begin{document}

\title[\xp]{\xp: An analysis package for autonomous gravitational-wave burst searches}
\author{Patrick J. Sutton}
\email{patrick.sutton@astro.cf.ac.uk}
\author{Gareth Jones}
\address{School of Physics and Astronomy, Cardiff University, Cardiff, United Kingdom, CF24 3AA}
\author{Shourov Chatterji}
\affiliation{Massachusetts Institute of Technology, Cambridge, MA 02139, USA}
\author{Peter Kalmus}
\affiliation{California Institute of Technology, Pasadena, CA  91125, USA}
\author{Isabel Leonor}
\affiliation{University of Oregon, Eugene, OR  97403, USA}
\author{Stephen Poprocki}
\affiliation{Department of Physics, Cornell University, Ithaca, NY 14853, USA}
\author{Jameson Rollins}
\affiliation{Columbia University, New York, NY  10027, USA}
\author{Antony Searle}
\affiliation{California Institute of Technology, Pasadena, CA  91125, USA}
\author{Leo Stein}
\affiliation{Massachusetts Institute of Technology, Cambridge, MA 02139, USA}
\author{Massimo Tinto}
\affiliation{Jet Propulsion Laboratory, California Institute of Technology, Pasadena, CA 91109}
\author{Michal Was}
\affiliation{LAL, Univ Paris-Sud, CNRS/IN2P3, Orsay, France}

\begin{abstract}
Autonomous gravitational-wave searches -- fully automated analyses 
of data that run without human intervention or assistance -- are 
desirable for a number of reasons. They are necessary for the rapid 
identification of gravitational-wave burst candidates, which in 
turn will allow for follow-up observations by other observatories
and the maximum exploitation of their scientific potential.  
A fully automated analysis would also 
circumvent the traditional ``by hand'' setup and tuning of 
burst searches that is both labourious and time consuming.  
We demonstrate a fully automated search with \xp, a 
software package for the coherent analysis of data from networks 
of interferometers for detecting bursts associated with GRBs and 
other astrophysical triggers.  We discuss the methods \xp~uses
for automated running, including background estimation, 
efficiency studies, unbiased optimal tuning of search thresholds, 
and prediction of upper limits.  These are all done automatically 
via Monte Carlo with multiple independent data samples, and without 
requiring human intervention.  As a demonstration of the power of 
this approach, we apply \xp~to LIGO data to compute the sensitivity to 
gravitational-wave emission associated with GRB 031108.
We find that \xp~is sensitive to signals approximately a factor 
of 2 weaker in amplitude than those detectable by the 
cross-correlation technique used in LIGO searches to date.
We conclude with the status of running \xp~as a 
fully autonomous, near real-time triggered burst search in 
the current LSC-Virgo Science Run. 
\end{abstract}


\maketitle

\section {Introduction}
\label{sec:intro}

Gravitational-wave bursts (GWBs) are one of the most interesting 
classes of signals being sought by the new generation of 
gravitational-wave detectors.  Possible sources include 
core-collapse supernovae \cite{Ott:2008wt}, the merger of binaries
containing black-holes or neutron-stars \cite{PhysRevLett.70.2984}, 
gamma-ray bursts \cite{Meszaros:2006rc}, and other relativistic 
systems; see \cite{CuTh:02} for a brief overview.
These systems typically involve matter at neutron-star 
densities and very strong gravitational fields, making GWBs potentially 
rich sources of information on relativistic astrophysics. 

The maximum exploitation of a GWB detection would occur when the system 
is observed by other ``messengers'' besides gravitational waves, such 
as in optical, gamma rays, or neutrinos \cite{Bloom:2009vx}.  Indeed, 
the first detection of a GWB might rely on independent confirmation by 
other observatories, and efforts are underway to develop collaborations 
between gravitational-wave detectors, electromagnetic telescopes, and 
neutrino observatories (see for example \cite{Kanner:2008zh,VanElewyck:2009pf}).
The rapid and confident identification of candidate GWBs by 
gravitational-wave detectors will be vital for these efforts. 

Unfortunately, the analysis of gravitational-wave data tends 
to be a slow process, with a typical latency of 
several years between the collection of the data and the 
publication of results.  For example, searches for 
gravitational-wave transients in the first year (2005-2006) 
of the LIGO Science Run 5 / Virgo Science Run 1 (\runone) have only 
recently been published \cite{S5firstyear,Abbott:2009tt}.
One of fastest such analyses has been the search for a 
gravitational-wave signal associated with GRB 070201 
\cite{Abbott:2007rh}, which was published 9 months after the event.

The rapid analysis of gravitational-wave data is not trivial, 
particularly given the non-stationary nature of the background 
noise in gravitational-wave detectors and the lack of accurate 
and comprehensive waveform models for GWB signals.
Specifically, we need methods capable of detecting weak signals 
with {\em a priori} unknown waveforms, yet which are 
simultaneously insensitive to the background noise ``glitches'' 
that are common in data from gravitational-wave detectors.
Glitch rejection is particularly important since it is the 
limiting factor in the sensitivity of current burst searches, and 
a confident detection of a GWB will depend critically on robust 
background estimation.  Detector characterisation 
\cite{Blackburn:2008ah,Leroy:2009zz} 
and search optimization tend to be laborious and time-consuming, 
as is accounting for other systematic effects such as uncertainties in 
detector calibration.

These considerations motivate the deployment of data analysis 
packages that can process data rapidly, yet comprehensively. 
The ideal scenario is a {\em fully autonomous} search -- one that runs 
continuously and without human intervention.  This requires an analysis 
that is self-tuning, adjusting search parameters to changes in the 
detector network and accounting for variations in the properties of the 
background noise around the time of candidate events.  

We present \xp~\cite{Xpipeline,Ch_etal:06}, a software package 
designed for autonomous searches for unmodelled gravitational-wave 
bursts.  \xp~targets GWBs associated 
with external astrophysical ``triggers'' such as gamma-ray bursts (GRBs), 
and has been used to search for GWBs associated with more 
than 100 GRBs that were observed during \runone~\cite{P0900023}.  
It performs a fully coherent analysis of data from arbitrary 
networks of gravitational-wave detectors, while being robust 
against noise-induced glitches.  
We emphasize the novel features of \xp, particularly 
a procedure for automated tuning of the background rejection tests. 
This allows the analysis of each external trigger to be optimized 
independently, based on background noise characteristics and detector 
performance at the time of the trigger, maximizing the search 
sensitivity and the chances of making a detection.  
This tuning uses independent data samples for tuning and estimating 
the significance of candidate events, for unbiased selection of GWB 
candidates. 
(See also \cite{0264-9381-25-10-105024} for a Bayesian-inspired 
technique for automated tuning.)
\xp~can also account automatically for effects like 
uncertainty in the sky position of astrophysical trigger and 
detector calibration uncertainties.
Furthermore, for the ongoing \runtwo~run, 
we are preparing the next step in the evolution of GWB 
searches: a fully autonomous search, wherein \xp~is triggered automatically 
by email reports of GRBs, and wherein 
data is analysed and candidate GWBs identified without human intervention.
Our goal is the complete analysis of each GRB within 24 hours of the receipt 
of the GRB notice.  Such a rapid analysis would be fast enough to allow 
further follow-up observations to be prompted by the GWB candidate.

We begin in Section~\ref{sec:theory} with a brief discussion of the 
theory of coherent analysis in gravitational-wave burst detection.  
In Section~\ref{sec:overview} we discuss the main steps followed in 
an \xp~triggered coherent search.  In Section~\ref{sec:grb} we 
demonstrate the sensitivity of \xp~on GRB 031108 using actual LIGO 
data, and compare to the upper limits set by the cross-correlation 
technique used in the published LIGO search for gravitational waves 
associated with the same GRB. 
In Section~\ref{sec:auto} we discuss the status of autonomous running 
of \xp~during the current S6-VSR2 science run of LIGO and Virgo. 
We conclude with a few brief comments in Section~\ref{sec:summary}.


\section{Coherent Analysis for Gravitational-Wave Burst Detection}
\label{sec:theory}

Most algorithms currently used in gravitational-wave burst detection 
can be grouped into two broad classes.  In {\em incoherent} methods 
\cite{AnBrCrFl:01,Sy:02}, candidate events typically are constructed 
from each detector data stream independently, and one looks for events 
with similar duration and frequency band that occur in all detectors 
simultaneously.  By contrast, {\em coherent} methods 
\cite{GuTi:89,Tinto:96,FlHu:98b,AnBrCrFl:01,Sy:03,Ca:04,WeSc:05,KlMoRaMi:05,
KlMoRaMi:06,MoRaKlMi:06,Ra:06,Ch_etal:06,Summerscales:2007xq,
Hayama:2007iz,Searle:2007uv,Searle:2008ap,Klimenko:2008fu} 
combine data from multiple detectors before processing, and create a 
single list of candidate events for the whole network. 
Coherent methods have some advantages over incoherent methods, such as   
demonstrated usefulness in rejecting background noise ``glitches'' 
\cite{Ca:04,WeSc:05,Ch_etal:06}, and for reconstructing GWB waveforms 
\cite{GuTi:89,Summerscales:2007xq}.
A less-recognized advantage of coherent methods is that they are 
relatively easy to tune.  For example, time-frequency coincidence 
windows for comparing candidate GWBs in different detectors are 
not necessary.  Detectors are naturally weighted  
by their relative sensitivity, so there is no need to tune the 
relative thresholds for generating candidate events in each detector.
This ease of tuning makes coherent methods particularly useful for 
rapid searches.

That said, there are also draw-backs to coherent methods, the most 
significant being computational cost.  Coherent combinations are 
typically a function of the sky position of the GWB source; there 
are $\gtrsim10^3$ resolvable directions on the sky for a worldwide 
detector network \cite{Fa:09}.  This cost is compounded by the need to estimate 
the background due to noise, which requires repeated re-analysis of 
the data using time shifts.  Fortunately, in triggered searches the 
sky position of the source is often known to high accuracy, and the 
amount of data to be analysed is relatively small (typically hours), 
so the computational cost of a fully coherent analysis is modest. 
This allows triggered searches to take advantage of the benefits of 
coherent methods while avoiding or minimizing most of the drawbacks.  

In this section we give a brief review of some 
of the main principles of coherent network analysis as implemented 
in \xp.


\subsection{Formulation}
\label{sec:notation}

A rigorous treatment of gravitational waves is based on linearized perturbations 
of the spacetime metric around a fixed background (see for example 
\cite{Flanagan:2005yc}).  In the linearized theory based on flat spacetime, when 
working in a suitable gauge, the perturbations representing the gravitational 
waves can be shown to obey the ordinary 
wave equation.  The gravitational waves are transverse, and 
travel at the speed of light. They have two independent polarizations, commonly 
referred to as ``plus'' ($+$) and ``cross'' ($\times$).  Their physical 
manifestation is a quadrupolar change in the distance between freely falling test 
particles (approximated in interferometric gravitational-wave detectors by the 
mirrors in the interferometer arms).  Explicit definitions of the plus and 
cross polarization states can be found, for example, in \cite{AnBrCrFl:01}.

The interferometers currently used to try to detect these waves are based on  
a laser, beamsplitter, and mirrors at the ends of each arm which serve as test masses.  
Data from each interferometer record the length difference of the arms and, 
when calibrated, measure the strain induced by a gravitational wave.  
The LIGO detectors are kilometer-scale power-recycled Michelson interferometers 
with orthogonal Fabry-Perot arms \citep{abbottnim04,abbott-2007}.  
There are two LIGO observatories: one located at Hanford, WA and the other at 
Livingston, LA.  The Hanford site houses two interferometers: one with 4 km 
arms (H1), and the other with 2 km arms (H2).  The Livingston observatory has 
one 4 km interferometer (L1).  The Virgo detector (V1) is in Cascina near Pisa, 
Italy.  It is a 3 km long power-recycled Michelson interferometer with orthogonal 
Fabry-Perot arms \citep{virgo08}.  
The GEO~600 detector \citep{geo08}, located near Hannover, Germany, is also 
operational, though with a lower sensitivity than LIGO and Virgo. 
These instruments are all designed to detect gravitational waves with frequencies ranging from 
$\sim30$~Hz to several kHz.  

Consider a gravitational wave $h_{+}(t,\vec{x}),h_{\times}(t,\vec{x})$ 
from a direction $\hat\Omega$.  The output of detector $\alpha\in[1,\ldots,D]$ 
is a linear combination of this signal and noise $n_\alpha$:
\bea\label{eqn:data}
	d_\alpha(t+\Delta t_\alpha(\hat\Omega))
		& = &  F_\alpha^+(\hat\Omega) h_+(t) + F_\alpha^\times(\hat\Omega) h_\times(t)
		\nonumber \\
		&   &  \mbox{}
		+ n_\alpha(t+\Delta t_\alpha(\hat\Omega)) \, .
\eea
Here $F^+(\hat\Omega)$, $F^\times(\hat\Omega)$ are the {\em antenna response 
functions} describing the sensitivity of the detector to the plus and cross 
polarizations (note that the choice of polarization basis is arbitrary; we 
use the $\psi=0$ choice of Appendix B of \cite{AnBrCrFl:01}).  Also, 
$\Delta t_\alpha(\hat\Omega)$ is the time delay 
between the position $\vec r_\alpha$ of detector $\alpha$ and an arbitrary 
reference position $\vec r_0$:
\be\label{eqn:delay}
	\Delta t_\alpha(\hat\Omega) = \frac{1}{c}(\vec r_0-\vec r_\alpha)\cdot\hat\Omega \, .
\ee
For brevity, we suppress explicit mention of the time delay and understand the 
data streams to be time-shifted by the appropriate amount prior to analysis.  
We also write $h_{+,\times}(t) \equiv h_{+,\times}(t,\vec{r}_0)$.  

Since the detector data is sampled discretely, we use discrete notation henceforth.
The discrete Fourier-transform $\tilde{x}[k]$ of a time-series $x[j]$ is
\bea\label{Fourier}
	\tilde{x}[k]  
		& = &  \sum_{j=0}^{N-1} x[j] \,\e^{-i2\pi jk/N} \, , \nonumber \\
	x[j]  
		& = &  \frac{1}{N} \sum_{k=0}^{N-1} \tilde{x}[k] \,\e^{i2\pi jk/N} \, ,
\eea
where $N$ is the number of data points in the time domain. 
Denoting the sampling rate by $f_s$, we can convert from continuous to
discrete notation using $x(t)\to x[j]$, $\tilde{x}(f)\to
f_s^{-1}\tilde{x}[k]$, $\int dt \to f_s^{-1}\sum_{j}$, $\int df \to
f_s N^{-1}\sum_{k}$, $\delta(t-t')\to f_s \delta_{jj'}$, and
$\delta(f-f')\to N f_s^{-1}\delta_{kk'}$.  
For example, the one-sided noise power spectral density 
$S_\alpha[k]$ of the noise $\tilde{n}_\alpha$ is
\be\label{eq:discstrainnoise}
	\langle \tilde{n}_\alpha^{*}[k] \tilde{n}_\beta[k'] \rangle
		=  \frac{N}{2} \delta_{\alpha\beta} \delta_{kk'} S_\alpha[k] \, ,
\ee
where the angle brackets indicate an average over noise instantiations.

It is conceptually convenient to define the noise-spectrum-weighted quantities  
\bea\label{eq:discwhiteningall}
	\tilde{d}_{\mathrm{w}\alpha}[k]
		& = & \frac{\tilde{d}_\alpha[k]}{\sqrt{\frac{N}{2}\psdk{\alpha}}}   \, , \\
	\tilde{n}_{\mathrm{w}\alpha}[k]
		& = & \frac{\tilde{n}_\alpha[k]}{\sqrt{\frac{N}{2}\psdk{\alpha}}}   \, , \\
	F^{+,\times}_{\mathrm{w}\alpha}(\hat\Omega,k) 
		& = & \frac{F^{+,\times}_\alpha(\hat\Omega)}{\sqrt{\frac{N}{2}\psdk{\alpha}}}   \, .
\eea
The normalization of the whitened noise is
\footnote{
More precisely, the $k$-dependent term on the right-hand side of 
(\ref{eq:discwhitening}) (the two-point spectral correlation function) 
is proportional to the Fourier transform of the window that was 
applied to the data before transforming to the frequency domain to 
suppress leakage.  We use $\delta_{kk'}$ in (\ref{eq:discwhitening}) 
as an approximation to simplify the notation.  
}
\be\label{eq:discwhitening}
\langle \tilde{n}_{\mathrm{w}\alpha}^*[k]\tilde{n}_{\mathrm{w}\beta}[k'] \rangle =
\delta_{\alpha\beta} \delta_{kk'} \, .
\ee
With this notation, equation~(\ref{eqn:data}) for the data 
measured from a set of $D$ detectors can be written in the simple 
matrix form
\be\label{eqn:matrix}
	\fdata = \F \fh + \fnoise \, ,
\ee
where we have dropped the explicit indices for frequency and sky position.
We use the boldface symbols 
$\fdata$, $\F$, $\fnoise$ to refer to noise-weighted quantities that 
are vectors or matrices on the space of detectors 
(note that $\fh$ is not noise-weighted and is not in the space of the detectors):
\be
	\fdata \equiv \left[
	\begin{array}{c}
		\tilde{d}_{\mathrm{w}1}\\ \tilde{d}_{\mathrm{w}2}\\ \vdots\\ \tilde{d}_{\mathrm{w}D}
	\end{array} \right] \, , \quad
	\fh \equiv  \left[
	\begin{array}{c}
			\tilde{h}_+ \\ \tilde{h}_\times
	\end{array} \right] \, , \quad
	\fnoise \equiv \left[
	\begin{array}{c}
		\tilde{n}_{\mathrm{w}1}\\ \tilde{n}_{\mathrm{w}1}\\ \vdots\\ \tilde{n}_{\mathrm{w}D}
	\end{array} \right] \, ,
\ee
and
\begin{eqnarray}\label{eqn:F}
\F & \equiv & \left[
	\begin{array}{c c}
		\Fplus & \Fcross  
	\end{array} \right]
	\equiv \left[
	\begin{array}{c c}
		F^+_{\mathrm{w}1} & F^\times_{\mathrm{w}1}  \\
		F^+_{\mathrm{w}2} & F^\times_{\mathrm{w}2} \\
		\vdots & \vdots\\
		F^+_{\mathrm{w}D} & F^\times_{\mathrm{w}D}
	\end{array} \right] 
	\, .
\end{eqnarray}
(See Table~\ref{tab:dimensions} for a list of the dimensions of all of the quantities 
used in this section.)
Note that each of these quantities is a function of both frequency and (through 
the antenna response or implied time shift) sky position.  As a consequence, 
coherent combinations typically have to be re-computed for every frequency bin 
as well as for every sky position.  Note also that, because 
of the noise-spectrum weighting, the whitened noise is isotropically distributed 
in the space of detectors [equation (\ref{eq:discwhitening})].  Therefore, 
all information on the sensitivity of the network both as a function of frequency 
and of sky position is contained in the matrix $\F$ defined by equation (\ref{eqn:F}).

\begin{table*}
\begin{center}
\begin{tabular}{|l|l|}
\hline
    quantity   &  dimensions \\
\hline
    $\h$, $\fh$, $\fhmax$  &  $2 \times 1$ vectors\\ 
	$\boldsymbol{F}$       &  $D \times 2$ matrix\\
	$\boldsymbol{P}^{\text{GW}}$, $\boldsymbol{P}^{\text{null}}$, $\boldsymbol{I}$  &  $D \times D$ matrices \\
	all other boldfaced symbols: $\data$, $\fdata$, $\Fplus$, $\FplusDP$, $\FplusHat$ etc.  &  $D \times 1$ vectors \\
\hline
\end{tabular}
\caption{\label{tab:dimensions} 
Dimensionality of various quantities used in this section. 
$D$ is the number of detectors in the network.
%
%
%
%
%
}
\end{center}
\end{table*}


\subsection{Standard Likelihood}
\label{sec:standard}

In this section we describe some of the simpler coherent likelihoods: 
those that can be computed from {\em projections} of the data.  These 
are the main ones used for signal detection in \xp.  We begin with the 
simplest coherent likelihood of all: the {\em standard} or 
{\em maximum likelihood}, first derived in \cite{FlHu:98b,AnBrCrFl:01}.

Let $P(\fdata | \fh)$ be the probability of obtaining the whitened data 
$\fdata$ in one time-frequency pixel in the presence of a known 
gravitational wave $\fh$ from a known direction.  Assuming Gaussian 
noise,
\begin{equation}
      \label{eqn:Ph}
	P(\fdata | \fh) =
		\frac{1}{(2\pi)^{D/2}}
		\exp\left[-\frac12\left|\fdata-\F\fh\right|^2\right].
\end{equation}
For a set $\{ \fdata \}$ of $N_\mathrm{p}$ time-frequency pixels,  
\begin{equation}
      \label{eqn:Phsum}
	P(\{ \fdata \} | \{ \fh \}) =
		\frac{1}{(2\pi)^{N_\mathrm{p} D/2}}
		\exp\left[-\frac12 \sum_k \left|\fdata[k]-\F[k]\fh[k]\right|^2\right] \, ,
\end{equation}
where $k$ indexes the pixels.  The likelihood ratio $L$ is defined by 
the log-ratio of this probability to the corresponding probability under the 
null hypothesis,
\begin{equation}
      \label{eqn:L}
	L\equiv\ln\frac{P(\{ \fdata \} | \{ \fh \})}{P(\{ \fdata \} | \{ 0 \} )} = 
		\frac12 \sum_k \left[ \left|\fdata\right|^2 - \left|\fdata-\F\fh\right|^2 \right] \, ,
\end{equation}
where $P(\{ \fdata \} | \{ 0 \} )$ is the probability of measuring the 
data $\{ \fdata \}$ when no GWB is present ($\fh=0$).

In practice, the signal waveform $\fh$ is not known {\em a priori}, 
so it is not clear how to compute the likelihood ratio~(\ref{eqn:L}). 
One approach is to treat the waveform values 
$\fh=(\tilde{h}_+,\tilde{h}_\times)$ in each time-frequency pixel 
as free parameters to be fit to the data.  The best-fit values 
$\fhmax$ are those that maximize the likelihood ratio:
\be\label{eqn:dL}
	0 = \left.\frac{\partial L}{\partial \fh}\right|_{\fh=\fhmax} \, .
\ee
Because the likelihood ratio $L$ is quadratic in $\fh$, (\ref{eqn:dL}) 
gives a linear equation for $\fhmax$.  The solution is
\be
	\fhmax = (\F^\ct\F)^{-1}\F^\ct \, \fdata \, ,
\ee
where we use $\mbox{}^\ct$ to denote the conjugate transpose.  
($\F$ is real, but other quantities such as the data vector 
$\fdata$ are complex.)
%
Substituting the solution for $\fhmax$ in (\ref{eqn:L}) gives the 
{\it standard likelihood},
\begin{equation}
      \label{eqn:SL}
	E_{\mathrm{SL}} \equiv 2 L(\fhmax) 
	= \sum_k \fdata^\ct \boldsymbol{P}^{\text{GW}} \fdata \, ,
\end{equation}
where we define
\be
	\boldsymbol{P}^{\text{GW}} \equiv \F \, (\F^\ct\F)^{-1}\F^\ct \, 
\ee
and we have used the fact that $\boldsymbol{P}^{\text{GW}}$ is Hermitian.
(The factor of 2 in the definition of $E_{\mathrm{SL}}$ is 
purely a matter of taste.)


\subsection{Projection Operators and the Null Energy}
\label{sec:projection}

It is easy to show that $\boldsymbol{P}^{\text{GW}}$ is a projection operator that projects 
the data into the subspace spanned by $\Fplus$ and $\Fcross$.
We know by equation~(\ref{eqn:data}) or (\ref{eqn:matrix})-(\ref{eqn:F}) that 
the contribution to $\fdata$ by any gravitational wave from a fixed sky position 
is restricted to this subspace.
The standard likelihood is therefore the maximum amount of 
energy \footnote{More precisely, since it is defined 
in terms of the noise-weighted data, the standard likelihood 
is the maximum possible squared signal-to-noise ratio $\rho^2$ that 
is consistent with the hypothesis of a gravitational wave from a given 
sky position.  See Section~\ref{sec:statistics}.} in the whitened data that is 
consistent with the hypothesis of 
a gravitational wave from a given sky position.

Contrast this with the {\it total energy} in the data, which is simply 
\begin{equation}
      \label{eqn:TOT}
	\Etot = \sum_k \left| \fdata \right|^2 \, .
\end{equation}
The total energy is an incoherent statistic in the sense that it 
contains only autocorrelation terms and no cross-correlation 
terms.  In the limit of a one-detector network, this is the quantity 
one computes for each time-frequency pixel in an excess-power 
search \cite{AnBrCrFl:01}.

The projection operator $\boldsymbol{P}^{\text{null}}\equiv(\boldsymbol{I}-\boldsymbol{P}^{\text{GW}})$, which is 
orthogonal to $\boldsymbol{P}^{\text{GW}}$, cancels the 
gravitational-wave signal.  This yields the {\em null stream} with 
energy
\begin{equation}
      \label{eqn:NULL}
	\En \equiv \Etot - \Esl 
		= \sum_k \fdata^\ct \boldsymbol{P}^{\text{null}} \fdata \, .
\end{equation}
The null energy is the minimum amount of energy in the whitened 
data that is inconsistent with the hypothesis of a gravitational wave 
from a given sky position.  

One advantage of coherent analysis is that the projection from 
the full data space with energy $\Etot$ to the subspace spanned 
by $\Fplus$, $\Fcross$ with energy $\Esl$ removes some fraction 
of the noise, with energy $\En$, without removing any of the 
signal component (small errors in calibrations, sky position, or 
power spectra change $\F$ but this affects the signal energy only 
at second order).  This means that a signal can be detected with 
higher confidence.  An important caveat is that the full benefit is 
gained only if the sky position is known {\em a priori}, such as in 
gamma-ray burst searches.  If the sky position of the source is 
not known {\em a priori}, one typically repeats the calculation of 
the likelihood for a set of directions spanning the entire sky 
($\gtrsim10^3)$ directions).  Since $\Fplus$, $\Fcross$ vary with the 
sky position, this means that many different projection operators 
will be applied to the data.  This will incur a false-alarm penalty.


\subsection{Dominant Polarization Frame and Other Likelihoods}
\label{sec:DPF}

For a single time-frequency pixel, the data from a set of $D$ 
detectors is a vector in a $D$-dimensional complex space.  
One basis of this space is formed by the set of single-detector 
strains (the basis in which all equations have been written thus 
far); however, this is not the most convenient basis for writing 
detection statistics. The $2$-dimensional subspace defined by 
$\Fplus,\,\Fcross$ is a natural starting point for the construction 
of a better basis.  If we examine the properties of this 
$2$-dimensional space, we find there is 
a direction (a choice of polarization angle) in which the detector 
network has the maximum antenna response, and an orthogonal
direction in which the network has minimum antenna response. 
Choosing those two directions as basis vectors, and completing 
them with an orthonormal basis for the null space, yields a very 
convenient basis in which to construct detection statistics. 
To further simplify things it is possible to define the $+,\,\times$ 
polarizations so that $\Fplus$ lies along the first basis vector, 
and $\Fcross$ along the second. This choice of polarization 
definition is called the \emph{dominant polarization frame} 
or DPF \cite{KlMoRaMi:05,KlMoRaMi:06}.
Note that while searches for modeled signals such as binary 
inspirals often select the polarization basis with reference 
to the source, the DPF polarization basis is tailored to the 
{\em detector network} at each frequency.  This makes it a 
particularly convenient choice when searching for more general 
gravitational-wave burst signals.

To see how one constructs the DPF, recall that the antenna 
response vectors in two frames separated by a polarization 
angle $\psi$ are related by
\begin{eqnarray}
      \label{eqn:Fpsi}
      \Fplus(\psi) & = & \cos2\psi \Fplus(0) +  \sin2\psi \Fcross(0) \, ,\\
      \Fcross(\psi) & = & -\sin2\psi \Fplus(0) +  \cos2\psi \Fcross(0) \, 
\end{eqnarray}
(see for example equations (B9), (B10) of \cite{AnBrCrFl:01}).
It is straightforward to show that for any direction on the sky, one 
can always chose a polarization frame such that $\Fplus(\psi)$ and 
$\Fcross(\psi)$ are orthogonal and $|\Fplus(\psi)| > |\Fcross(\psi)|$. 
Explicitly, given $\Fplus(0)$, $\Fcross(0)$ in the original polarization 
frame, the rotation angle $\psi_\mathrm{DP}$ giving the dominant 
polarization frame is
\begin{eqnarray}\label{eqn:psiDP}
\psi_{DP}(\hat\Omega,k)  
	& = &  \frac{1}{4} \mathrm{atan2}\left(2 \Fplus(0) \cdot \Fcross(0), 
	       \right .\nonumber \\
	&	&  \left. \quad
			|\Fplus(0)|^2 - |\Fcross(0)|^2\right) \, .
\end{eqnarray}
where $\mathrm{atan2}(y,x)$ is the arctangent function with range $(-\pi,\pi]$. 
Note that $\psi_{DP}$ is a function of both sky position 
and frequency (through the noise weighting of $\Fplus$ and $\Fcross$).

\begin{figure}
\begin{center}
  \includegraphics[width=0.4\textwidth]{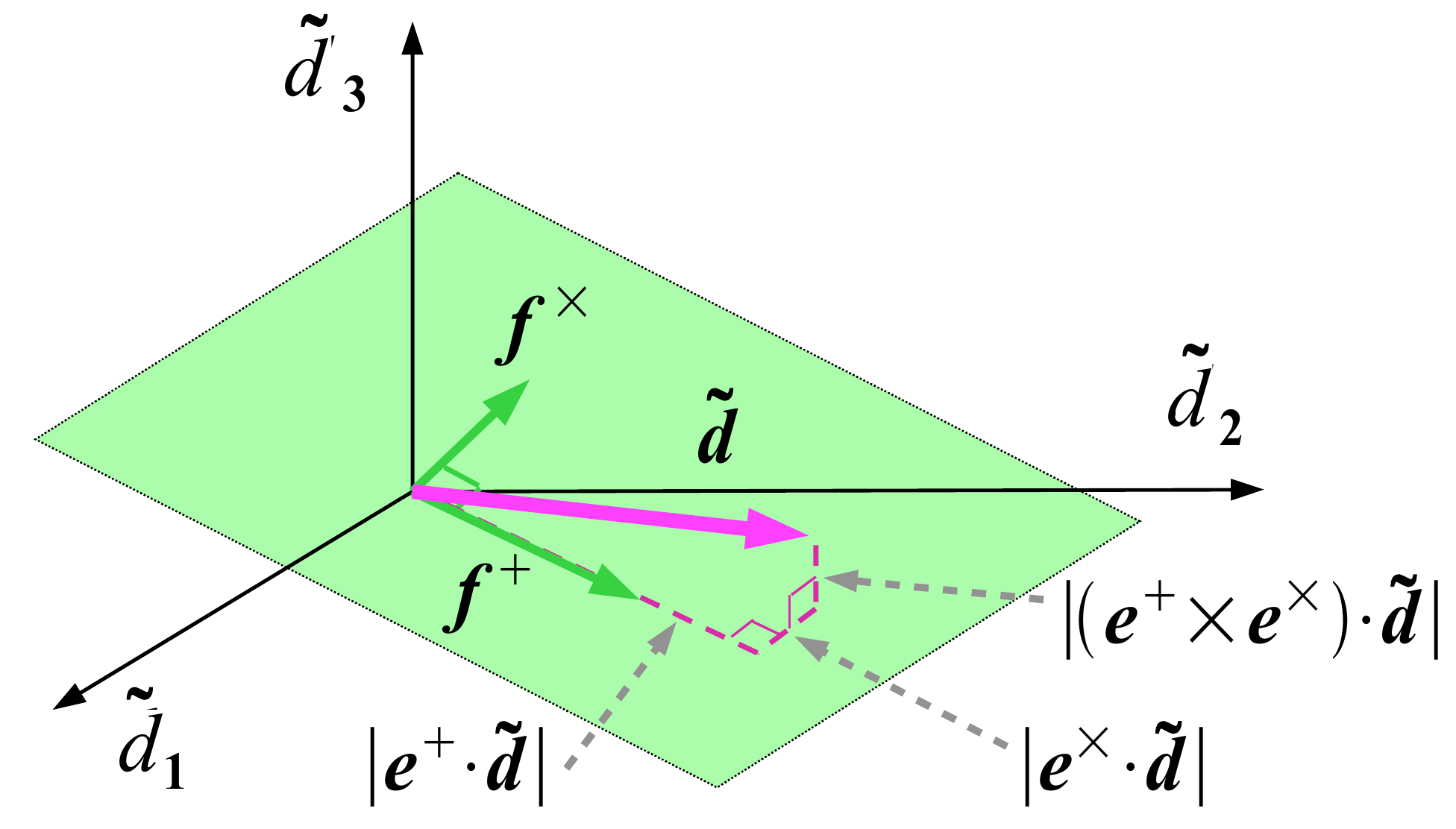}
  \caption{\label{fig:geometry}  
Space of detector strains for the 3-detector case for one data sample. 
The green plane is the plane spanned by the antenna response vectors 
$\FplusDP$, $\FplusDP$. 
The thick magenta line is the vector of detector strains $\fdata$ for one 
realization of noise and signal. 
The dashed lines show the projection of the data vector 
into the detector response plane and into the null space.
}
\end{center}
\end{figure}

We denote the antenna response vectors in the DPF by the lower-case 
symbols $\FplusDP$, $\FcrossDP$.  They have the properties
\begin{equation}
|\FplusDP|^2  \ge  |\FcrossDP|^2 \, , 
\end{equation}
\begin{equation}
\FplusDP \cdot \FcrossDP  =  0 \, .
\end{equation}
In the DPF the unit vectors $\FplusHat\equiv\FplusDP/|\FplusDP|$, 
$\FcrossHat\equiv\FcrossDP/|\FcrossDP|$ are part of an orthonormal 
coordinate system; see Figure~\ref{fig:geometry}.  
Indeed, the DPF can be viewed as the natural 
coordinate system in the space of detector data for understanding 
the sensitivity of the network.  Mathematically, rotating to the DPF 
is the same as doing a {\em singular value decomposition} of the 
matrix $\F$.  The singular values are $|\FplusDP|^2$ and 
$|\FcrossDP|^2$; i.e., the magnitudes of the antenna response 
evaluated in the DPF.

It should be noted that the DPF does not specify any particular 
choice of basis for the null space.  Convenient choices for the 
null basis can be motivated by how the null energy is used in the 
search, but we do not consider this issue here. 

In the DPF, the projection operator $\boldsymbol{P}^{\text{GW}}$ takes on the very simple form 
\begin{equation}\label{eqn:ProjDP}
	\boldsymbol{P}^{\text{GW}} = \FplusHat \FplusHat{}^\ct + \FcrossHat \FcrossHat{}^\ct ~ .
\end{equation}
The standard likelihood (\ref{eqn:SL}) becomes
\begin{equation}\label{eqn:SLDPF}
	\Esl = \sum_k \left[ \left| \FplusHat \cdot \fdata \right|^2 
	      + \left| \FcrossHat \cdot \fdata \right|^2 \right] \, ,
\end{equation}
where we use the notation $\boldsymbol{a}\cdot\boldsymbol{b}$ to denote the familiar 
dot product between $D \times 1$ dimensional vectors $\boldsymbol{a}$ and $\boldsymbol{b}$.
The {\em plus energy} or {\em hard constraint likelihood} 
\cite{KlMoRaMi:05,KlMoRaMi:06} is the energy in the $\hp$ polarization 
in the DPF:
\begin{equation}\label{eqn:HC}
	\Ep \equiv \sum_k \left| \FplusHat \cdot \fdata \right|^2 \, .
\end{equation}
The {\em cross energy} is defined analogously:
\begin{equation}
	\Ec \equiv \sum_k \left| \FcrossHat \cdot \fdata \right|^2 \, .
\end{equation}
The {\em soft constraint likelihood} \cite{KlMoRaMi:05,KlMoRaMi:06} (not a projection likelihood) is
\begin{equation}
      \label{eqn:SC}
	\Esc \equiv \sum_k \left[ \left| \FplusHat \cdot \fdata \right|^2 
			+ \epsilon \left| \FcrossHat \cdot \fdata \right|^2 \right] \, ,
\end{equation}
where the weighting factor $\epsilon$ is defined in the DPF as 
\begin{equation}
      \label{eqn:epsilon}
	\epsilon \equiv \frac{\left| \FcrossDP \right|^2}{\left| \FplusDP \right|^2} \in [0,1] \, .
\end{equation}
Typical values are $\epsilon \sim 0.01 - 0.1$ for the LIGO network.

Numerous other likelihood-based coherent statistics have been 
introduced in the literature, such as the Tikhonov regularized 
statistic \cite{Ra:06}, a sky-map variability statistic 
\cite{Hayama:2007iz}, and modified constraint likelihood 
statistics \cite{Klimenko:2008fu}.  
Also, comprehensive Bayesian formulations of the problem of GWB 
detection and waveform estimation are described in 
\cite{Summerscales:2007xq,Searle:2007uv,Searle:2008ap}.
While some of these statistics are available in \xp, we do not 
consider them here.


\subsection{Statistical Properties}
\label{sec:statistics}

One convenient property of the projection likelihoods $\Ep$, $\Ec$, 
$\Esl$, $\En$, $\Etot$ is that their statistical properties for signals 
in Gaussian background noise are very simple.  Specifically, for a 
set of time-frequency pixels and a sky position chosen {\em a priori}, 
each of these energies follows a $\chi^2$ distribution with 
$2N_\mathrm{p}D_\mathrm{proj}$ degrees of freedom:
\be\label{eqn:chi2}
2E \sim \chi^2_{2N_\mathrm{p}D_\mathrm{proj}}(\lambda) \, .
\ee
Here $N_\mathrm{p}$ is the number of pixels (or time-frequency 
volume). $D_\mathrm{proj}$ is the number of dimensions of 
the projection, which is 1 for $\Ep, \Ec$, 2 for $\Esl$, and $D$ for $\Etot$.  
$D_\mathrm{proj} = D-2$ for $\En$, except when the null stream 
is constructed as the difference of the data streams from the two 
co-aligned LIGO-Hanford detectors, H1 and H2, in which case it is $D-1$ 
(the H1-H2 sub-network is only sensitive to a single gravitational-wave 
polarization, so only one dimension is removed in forming the null stream).  
The factor of 2 in the degrees of freedom 
occurs because the data are complex.
The non-centrality parameter $\lambda$ 
is the expected squared signal-to-noise ratio of a matched filter 
for the waveform restricted to the time-frequency region in 
question \footnote{Equations (\ref{eqn:rhop})--(\ref{eqn:rhotot}) assume that the 
sum is restricted to positive frequencies; if negative frequencies 
are included then the non-centrality parameters should be doubled 
($\lambda\to2\rho$).  In both cases $\rho$ is to be understood as the 
expected output of a matched filter.} and after projection by the appropriate likelihood 
projection operator, summed over the network:
\bea
\lambda_+ & = & 2 \sum_k |\FplusDP|^2 |\hp|^2 
	\nonumber \\[-3mm]
	& = &  \frac{4}{N} \sum_\alpha \sum_k 
		\frac{\left| F^+_\alpha(\psi_{DP}) \tilde{h}_+[k](\psi_{DP}) \right|^2}{S_\alpha[k]}
	=:  \rho_+^2 , \label{eqn:rhop} \\
\lambda_\times & = & 2 \sum_k |\FcrossDP|^2 |\hc|^2
	\nonumber \\[-3mm]
	& = &   
	    \frac{4}{N} \sum_\alpha \sum_k 
		\frac{\left| F^\times_\alpha(\psi_{DP}) \tilde{h}_\times[k](\psi_{DP}) \right|^2}{S_\alpha[k]}
	=:  \rho_\times^2 ,  \qquad \label{eqn:rhoc} \\
\lambda_\mathrm{SL} & = & 2 \sum_k 
		\left[ |\FplusDP|^2 |\hp|^2 + |\FcrossDP|^2 |\hc|^2 \right]
		\nonumber \\[-3mm]
	& = &  \frac{4}{N} \sum_\alpha \sum_k \frac{\left| F^+_\alpha \tilde{h}_+[k] 
		+ F^\times_\alpha \tilde{h}_\times[k] \right|^2}{S_\alpha[k]}
	=:  \rho^2 \, , \label{eqn:rho} \\
\lambda_\mathrm{tot}\!\! & = & \rho^2 \, , \label{eqn:rhotot} \\
\lambda_\mathrm{null}\!\!\! & = & 0 \, .
\eea
Note that in (\ref{eqn:rhop}) and (\ref{eqn:rhoc}) the antenna responses and 
waveforms are defined in the DPF.  Eqn.~(\ref{eqn:rho}) is actually independent 
of the polarization basis used.

The mean and standard deviation of the non-central $\chi^2$ 
distribution~(\ref{eqn:chi2}) are 
$(2N_\mathrm{p}D_\mathrm{proj} + \lambda)$ and 
$\sqrt{2N_\mathrm{p}D_\mathrm{proj}}$.  Consequently, one 
expects a signal to be detectable by a given coherent statistic 
when 
\be\label{eqn:detection}
\frac{\lambda}{\sqrt{2N_\mathrm{p}D_\mathrm{proj}}} \gg 1 \, .
\ee
Table~\ref{tab:likelihoodVariations} shows the mean and standard 
deviation of various energy measures when the correct sky position 
and the time-frequency region are known {\em a priori}.  

For a circularly polarized or unpolarized gravitational wave, 
$\rho_\times^2/\rho_+^2 \simeq \epsilon \ll 1$ for 
typical sky positions.  
For example, for the LIGO-Virgo network of detectors H1-H2-L1-V1, 
assuming H2 is half as sensitive as H1, L1, and V1, the median 
value of $\epsilon$ is 0.1, while for the LIGO network H1-H2-L1 
the median is 0.02.
%
%
As a consequence, for many signals 
$\rho_\times^2$ is negligible.  (An exception is linearly polarized 
GWBs; for these the random polarization angle can make 
$\rho_\times^2>\rho_+^2$ in the H1-H2-L1 network for approximately 
10\% of signals for a typical sky position.)  Since all of the energies 
except $\Ec$ in Table~\ref{tab:likelihoodVariations} include 
$\rho_+^2$, their relative performance is dominated by the 
level of noise fluctuations.  The noise fluctuations in the energies 
scale as the square-root of the number of orthogonal directions 
used to compute the energy.  As a consequence, we expect those 
statistics that project the data down to fewer dimensions to perform 
better for GWB detection.  For $\Ep$ the data is projected onto 
a single direction.  $\Esl$ and $\Esc$ use data along two 
directions, and so have higher noise.  The total energy $\Etot$ 
uses all of the data and therefore incorporates the largest 
contributions from noise.  In practice, coherent consistency tests
(discussed in the next section) can be used to reduce the noise 
background, allowing statistics like $\Esl$ to be used effectively, 
so that all of the signal-to-noise ratio of a GWB ($\rho_+^2$ and 
$\rho_\times^2$) can be included in the detection statistic.

\begin{table*}
\begin{center}
\begin{tabular}{|r|r|r|}
\hline
	Energy measure &
	$\mathrm{mean}(2E)|_\mathrm{GWB}$ &
	$\displaystyle{\frac{\mathrm{std}(2E)|_\mathrm{noise}}{\sqrt{2N_\mathrm{p}}}}$\\
\hline
	$E_{\mathrm{TOT}}$ &
	$\rho^2 = \rho_+^2 + \rho_\times^2$ &
	$\sqrt{D}$\\
$E_{\mathrm{SL}}$ & 
	$\rho^2 = \rho_+^2 + \rho_\times^2$ &
	$\sqrt{2}$\\
$E_{\mathrm{+}}$ & 
	$\rho_+^2$ &
	1\\
$E_{\mathrm{\times}}$ & 
	$\rho_\times^2$ &
	1\\
$E_{\mathrm{soft}}$ &
	$\simeq(\rho_+^2 + \epsilon\rho_\times^2)$ &
	$\sqrt{1+\epsilon}$ \\
\hline
\end{tabular}
\caption{\label{tab:likelihoodVariations} 
Expected signal and noise contributions to various coherent energies.
The mean($2E$) column shows the contribution to the mean energy
due to a gravitational wave, evaluated in the dominant polarization frame.
See equations (\ref{eqn:epsilon}), (\ref{eqn:rhop})--(\ref{eqn:rhotot}).
The std($2E$) column shows the standard deviation due to noise 
fluctuations assuming a non-aligned detector network ({\em i.e.\/}, $|\FcrossDP|>0$).  
The values for $\Esc$ are written as approximate because the 
weighting factor $\epsilon$ is itself a function of frequency.
All of these values assume the time-frequency region to sum over 
and the correct sky location are known {\em a priori}.
}
\end{center}
\end{table*}


\subsection{Incoherent Energies and Background Rejection}
\label{sec:incoherent}

The various likelihood measures $\Esl$, $\Ep$, {\em etc.} are 
motivated as detection statistics under the assumption of stationary 
Gaussian background noise.  Real detectors do not have purely 
Gaussian noise.  Rather, real detector noise contains \emph{glitches}, 
which are short transients of excess strain that can masquerade as 
gravitational-wave burst signals.  In practice, without a means to 
distinguish noise glitches from true GW signals, the sensitivity 
of a burst search will be limited by such glitches.  Coherent analyses 
can be particularly susceptible to such false alarms, since even a 
glitch in a single detector will produce large values for likelihoods such 
as $\Esl$.  In this section we outline a technique for the effective 
suppression of such false alarms in coherent analyses.

As shown in Chatterji {\em et al.}  \cite{Ch_etal:06}, one can use the 
autocorrelation component of coherent energies to construct tests 
that are effective at rejecting glitches.  This coherent veto test is 
based on the null space -- the subspace orthogonal to that used to 
define the standard likelihood.  The projection of $\data$ on this 
subspace contains only noise, and the presence or absence of GWs 
should not affect this projection in any way.  By contrast, glitches do 
not couple into the data streams with any particular relationship to 
$\Fplus, \Fcross$.  As a result, glitches will generally be present in 
the null space projection.  This provides a way to distinguish true 
GWs from glitches, by requiring the null energy to be small for a 
transient to be considered a GW~\cite{WeSc:05}.

To see how an effective test can be constructed, note that we can 
write equation (\ref{eqn:NULL}) for the null energy as
\begin{eqnarray}
 \En & = & \sum_k \sum_{\alpha,\beta} \tilde{d}^\ct_\alpha 
		P^\mathrm{null}_{\alpha \beta} \tilde{d}^{\vphantom{\ct}}_\beta \, .
\end{eqnarray}
%
As pointed out in Chatterji {\em et al.}  \cite{Ch_etal:06}, the null 
energy is composed of cross-correlation terms 
$\tilde{d}^\ct_{\alpha} \tilde{d}^{\vphantom{\ct}}_\beta$ and 
auto-correlation terms $\tilde{d}^\ct_\alpha \tilde{d}^{\vphantom{\ct}}_\alpha$.
If the transient signal is not correlated between detectors 
(as is expected for glitches), then the cross-correlation terms
will be small compared to the auto-correlation terms.  As a 
consequence, for a glitch we expect the null energy to be 
dominated by the auto-correlation components:
\begin{equation}
 \En  \simeq  \In  \equiv \sum_k \sum_{\alpha} 
		P^\mathrm{null}_{\alpha\alpha} |\tilde{d}_\alpha|^2  \, .
		\qquad \mathrm{(glitches)}
\end{equation}
This auto-correlation part of the null energy is called the 
\emph{incoherent energy}. 

By contrast, for a GW signal, the transient is correlated between
the detectors according to equations (\ref{eqn:data}) or  
(\ref{eqn:matrix})--(\ref{eqn:F}).  By construction of the null 
projection operator, these correlations cancel in the null stream, 
leaving only Gaussian noise.  They cannot cancel in $\In$, 
however, since that is a purely incoherent statistic.  Therefore, 
for a strong GW signal we expect
\begin{equation}
	\En  \ll  \In \, . \qquad \mathrm{(GW)}
\end{equation}
Based on these considerations, the coherent veto test introduced 
by Chatterji {\em et al.} \cite{Ch_etal:06} is to keep only transients 
with 
\begin{equation} \label{ratioC}
	\In/\En > C, 
\end{equation}
where $C$ is some constant greater than 1.  This test 
is particularly effective at eliminating large-amplitude glitches. 
For smaller amplitude glitches $\En$ can be small compared to $\In$ 
due to statistical fluctuations; for this reason in \xp\ we use a 
modified test where the effective threshold $C$ varies with the event 
energy, as discussed in Section~\ref{sec:veto}.

Analogous tests can be imposed on the other coherent energies, 
$\Ep$, $\Ec$, {\em etc.}.  We define the corresponding incoherent 
energies by 
\begin{eqnarray}
\Ip  & \equiv &  \sum_k \sum_{\alpha} 
		\left|e^+_\alpha \tilde{d}_\alpha\right|^2  \, , \\
\Ic  & \equiv &  \sum_k \sum_{\alpha} 
		\left|e^\times_\alpha \tilde{d}_\alpha\right|^2  \, , \\
\Isl  & \equiv &  \sum_k \sum_{\alpha} \left[
		\left|e^+_\alpha \tilde{d}_\alpha\right|^2 
		+ \left|e^\times_\alpha \tilde{d}_\alpha\right|^2 \right] 
		= \Ip + \Ic \, . \qquad
\end{eqnarray}
In each case, we compare the coherent energy $E$ to its incoherent 
counterpart $I$, making use of the expectation that for a glitch, 
$E \simeq I$.  For a strong GW, the signal summed over both 
polarizations should build coherently, so one will find 
\be
	\Esl > \Isl  \, . \qquad \mathrm{(GW)}
\ee
By contrast, one may find $\Ep > \Ip$ or $\Ep < \Ip$ depending on 
the polarization of the GW signal.  Specifically, if the GW signal is 
predominantly in the $+$ polarization in the DPF, then one will find 
\be
	\begin{array}{c} \Ep > \Ip \\ \Ec < \Ic \end{array} \qquad (\mathrm{signal~predominantly~}\hp) \, .
\ee
If the GW signal is predominantly in the $\times$ polarization 
in the DPF, then one will find the reverse:
\be
	\begin{array}{c} \Ep < \Ip \\ \Ec > \Ic \end{array} \qquad (\mathrm{signal~predominantly~}\hc) \, .
\ee
In general, a GW will be characterised by at least one of 
$\Ep>\Ip$ or $\Ec>\Ic$; {\em i.e.}, at least one of the 
polarizations will show a coherent buildup of signal-to-noise
across detectors.  This allows us to impose coherent 
glitch rejection tests even in the case where a null stream is 
not available, such as the H1-L1 network of LIGO detectors.  
Specific examples of coherent consistency tests are discussed in 
Sections~\ref{sec:veto} and~\ref{sec:grb}.

These incoherent energies are not defined as the magnitude of 
a projection.  As a result, they do not obey $\chi^2$ statistics.  
They do, however, obey a simple relation with the coherent energies:
\be
	\Ip + \Ic + \In = \Ep + \Ec + \En = \Etot \, .
\ee
Equivalently, the sum of the cross-correlation contributions to 
$\Ep$, $\Ec$, and $\En$ cancel:
\bea
0 & = & (\Ep-\Ip) + (\Ec-\Ic) + (\En-\In) ~ \nonumber \\
  & = & (\Esl-\Isl) + (\En-\In)  \, .
\eea

\section{Overview of {\xp}}
\label{sec:overview}

\xp~is a \matlab-based software package for performing coherent searches 
for gravitational-wave bursts in data from arbitrary networks of detectors.
In this section we give an overview of the main steps followed in a triggered 
burst search, describing how the data is processed and how candidate GWBs are 
identified.  In Section~\ref{sec:auto} we discuss how an \xp~analysis is triggered.

\subsection{Preliminaries}
\label{sec:prelim}

\xp~performs the coherent analyses described in Section~\ref{sec:theory}.
The user (a human or automated triggering software) specifies:
\begin{enumerate}
\item
a set of detectors;
\item
one or more intervals of data to be analysed;
\item
a set of coherent energies to compute;  
\item
a set of sky positions; and 
\item
a list of parameters (such as FFT lengths) for the analysis.
\end{enumerate}
In standard usage, \xp~processes the data and produces lists of candidate 
gravitational-wave signals for each of the specified sky positions.  
It does this by first constructing time-frequency maps of the various 
energies in the reconstructed $\hp$, $\hc$, and null streams.  
\xp~then identifies clusters of pixels with large values of one of the 
coherent energies, such as $\Esl$ or $\Ep$.

\subsection{Time-frequency maps}
\label{sec:tfmaps}

\begin{figure}[htb]
\begin{center}
  \includegraphics[width=0.5\textwidth]{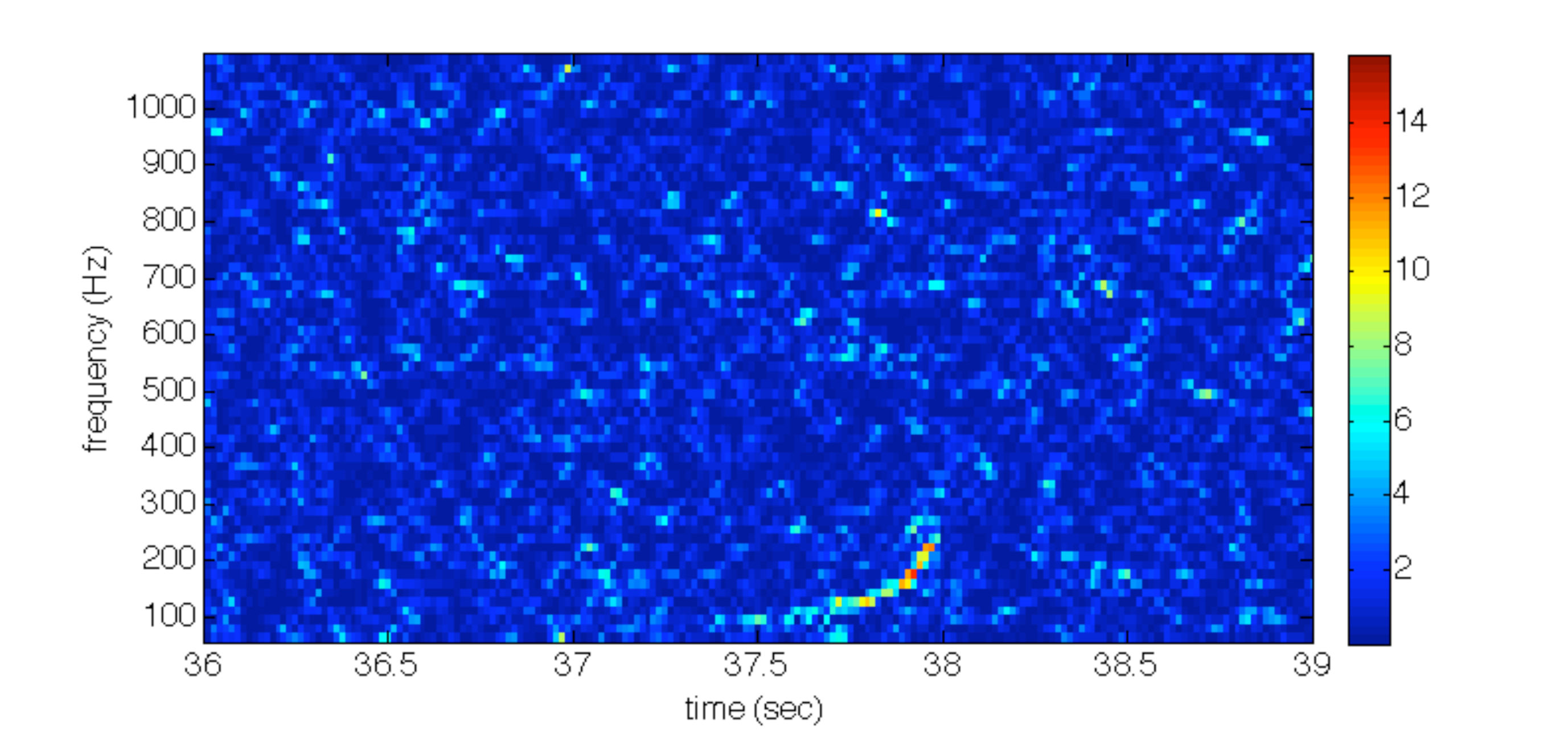}
  \includegraphics[width=0.5\textwidth]{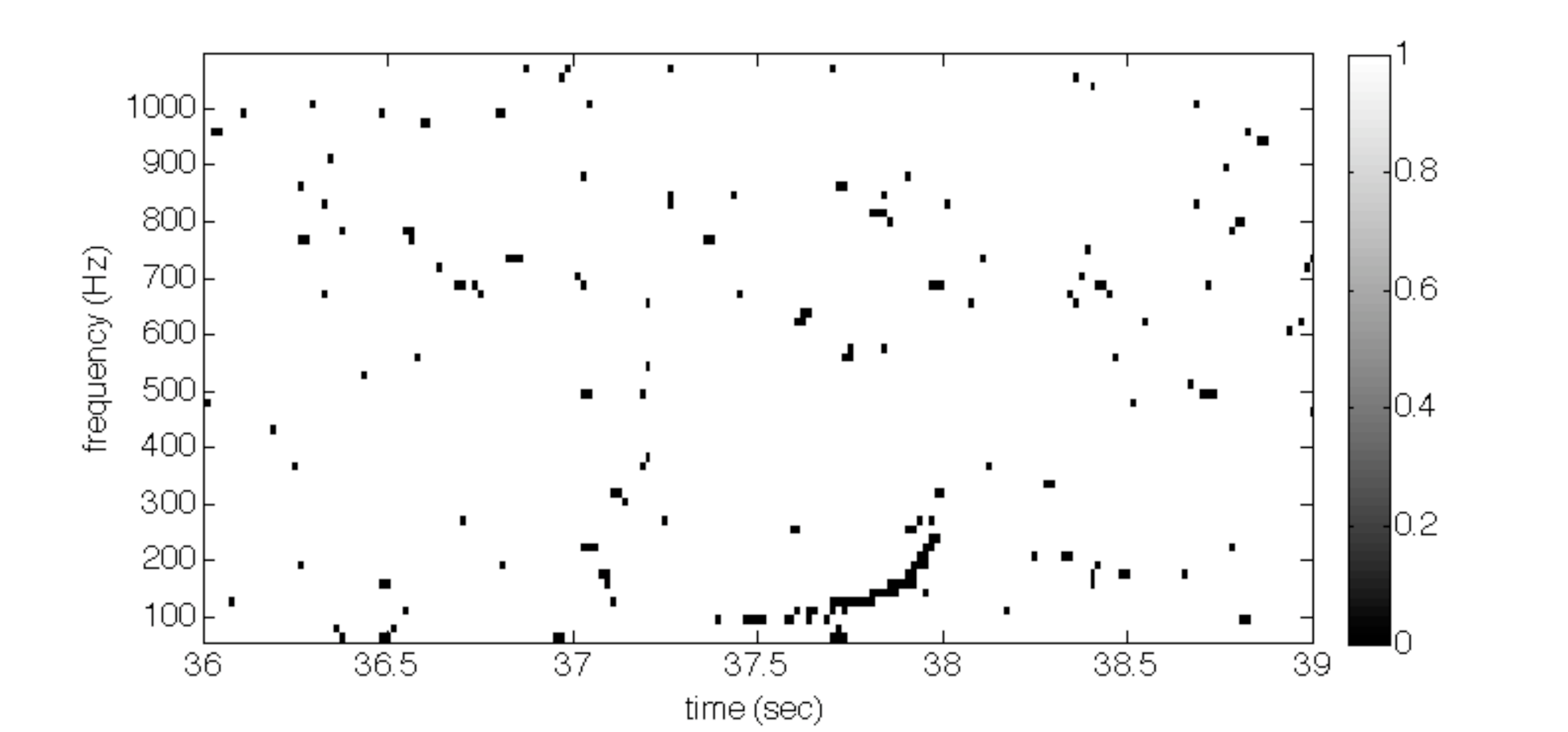}
  \caption{\label{fig:maps} 
A simulated $1.4 M_\odot -10.0 M_\odot$ neutron star Ð black hole inspiral 
at an effective distance of 37 Mpc, added to simulated noise from the two 
LIGO-Hanford detectors. 
(top) Time-frequency map of the $\Ep$ energy.
(bottom) The highest 1\% of pixels highlighted. The inspiral ``chirp'' is clearly visible.}
\end{center}
\end{figure}

\xp~typically processes data in 256 s blocks.  
First, it loads the requested data.  It constructs a zero-phase linear predictor error 
filter to whiten the data and estimate the power spectrum \cite{Ch_etal:06,ChBlMaKa:04}.  
For each sky position, 
\xp~time-shifts the data from each detector according to equations~(\ref{eqn:data}) and  
(\ref{eqn:delay}).  The data is divided into overlapping segments and Fourier 
transformed, producing time-frequency maps for each detector.   
Given the time-frequency maps for the individual detector data streams $\fdata$, 
\xp~coherently sums and squares these maps in each pixel to produce time-frequency maps 
of the desired coherent energies; see Figure~\ref{fig:maps}.  This representation 
gives easy access to the temporal evolution of the spectral properties of the 
signal, and all statistics and other quantities that are functions of time and frequency.

\subsection{Clustering and Event Identification}
\label{sec:clustering}

Given time-frequency maps of each of the coherent energies, the 
challenge is then to identify potential gravitational-wave signals in 
these maps.  

\begin{figure}
\begin{center}
  \includegraphics[width=0.2\textwidth]{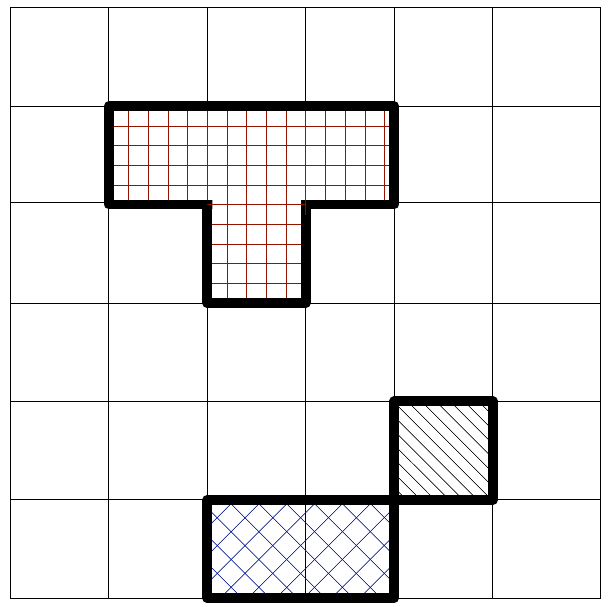}
  \caption{\label{fig:clusters}
A time frequency map with ``black'' pixels grouped into 3 clusters.  
Nearest-neighbor black pixels (those that share an edge) are 
grouped into a single cluster.  Each cluster in this 
image is denoted by a different color and hatching pattern.
}
\end{center}
\end{figure}

The approach used in \xp~is pixel clustering \cite{Sy:02}.  The user 
singles out one of the energy measures -- typically $\Esl$, 
the summed energy in the reconstructed $\hp$ and $\hc$ streams -- 
as the {\em detection statistic}.  
A threshold is applied to the detection statistic map so that a fixed 
percentage ({\em e.g.}, $1\%$) of the pixels with the highest value 
in the current map are marked as \emph{black pixels}; see Figure~\ref{fig:maps}.  
Following the method of \cite{Sy:02}, black pixels that share a common 
side (nearest neighbors) are grouped together into clusters; see 
Figure~\ref{fig:clusters} for an example.  (As allowed in \cite{Sy:02}, 
the user may specify a different connectivity criterion, such as next-nearest 
neighbors, or apply the ``generalized clustering'' procedure.)
This clustering technique is appropriate for a GWB whose shape 
in the time-frequency plane is connected, as opposed to consisting 
of well-separated ``blobs''.  This assumption is valid for many 
well-modeled signals such as low-mass inspirals and ringdowns. 

Each cluster is considered a candidate detection event.   
Each is assigned a detection statistic value from its constituent pixels by 
simply summing the values of the statistic in the pixels. This is motivated by the 
additive property of the log-likelihood ratio -- the inherited detection statistic is 
exactly the detection statistic for the area defined by the cluster.  Each cluster 
is also assigned an approximate statistical significance ${\cal S}$ based on the 
$\chi^2$ distribution; see equation (\ref{eqn:chi2}).  This significance is used 
when comparing different clusters to determine which is the ``loudest'' -- the 
best candidate for being a gravitational wave signal.
Finally, the energy at the same time-frequency locations in maps of each of the other 
requested likelihoods is also computed and recorded for each cluster.  

The clusters are saved for later post-processing.  The analysis 
of time shifting (see equation~(\ref{eqn:delay})), FFTing, and cluster identification is then repeated for 
each of the other sky positions and for each of the requested FFT 
lengths.

One other important feature of the time-frequency maps is the Fourier 
transform length or {\em analysis time} $T$, which determines the 
aspect ratio of the pixels. A longer time gives pixels with poor time 
resolution but good frequency resolution; a shorter time gives pixels 
with good time resolution but poor frequency resolution. Depending on 
the signal duration, 
different analysis times may be optimal. Since each pixel has the same 
noise distribution (assuming Gaussian statistics), the optimal pixel 
size is the size for which the signal spans the smallest number of 
pixels, so that the statistic is least polluted by noise.

Since the optimal analysis time for the incoming signal is not known, 
\xp~uses several analysis times, and applies a second layer of clustering 
between analysis times. For this second layer of clustering, clusters made 
from black pixels at two different analysis times that overlap in time and 
frequency are compared. The cluster that has the largest significance is 
kept as a candidate event; the less significant overlapping clusters are 
discarded.

\subsection{Glitch rejection}
\label{sec:veto}

As noted in Section~\ref{sec:theory}, noise glitches tend to have a strong 
correlation between each coherent energy $\En$, $\Ep$, $\Ec$, and its 
corresponding incoherent energy $\In$, $\Ip$, $\Ic$.  \xp~compares the 
coherent and incoherent energies to veto events that have properties similar 
to the noise background.  
These {\em coherent veto tests} are applied in post-processing ({\em i.e.}, 
after candidate events from the different analysis times are generated and 
combined).

Two types of coherent veto are available in \xp.  Both are pass/fail tests.  
The simplest is a threshold on the ratio $I/E$.  Following the discussion 
in Section~\ref{sec:incoherent}, a cluster passes the coherent test if 
\bea
\In / \En & \ge & r_\mathrm{null} \, , \label{eqn:ration} \\
|\log_{10}(\Ip / \Ep)| & \ge & \log_{10}(r_+) \, , \label{eqn:ratiop} \\
|\log_{10}(\Ic / \Ec)| & \ge & \log_{10}(r_\times) \, , \label{eqn:ratioc}
\eea
where the thresholds $r_\mathrm{null}$, $r_+$, and $r_\times$ may be specified 
by the user or chosen automatically by \xp.  The form of equations 
(\ref{eqn:ratiop}) and (\ref{eqn:ratioc}) make these tests two-sided; 
{\em i.e.}, they pass clusters that are sufficiently far above {\em or} 
below the diagonal.

\begin{figure}[htb]
\begin{center}
  \includegraphics[width=0.45\textwidth]{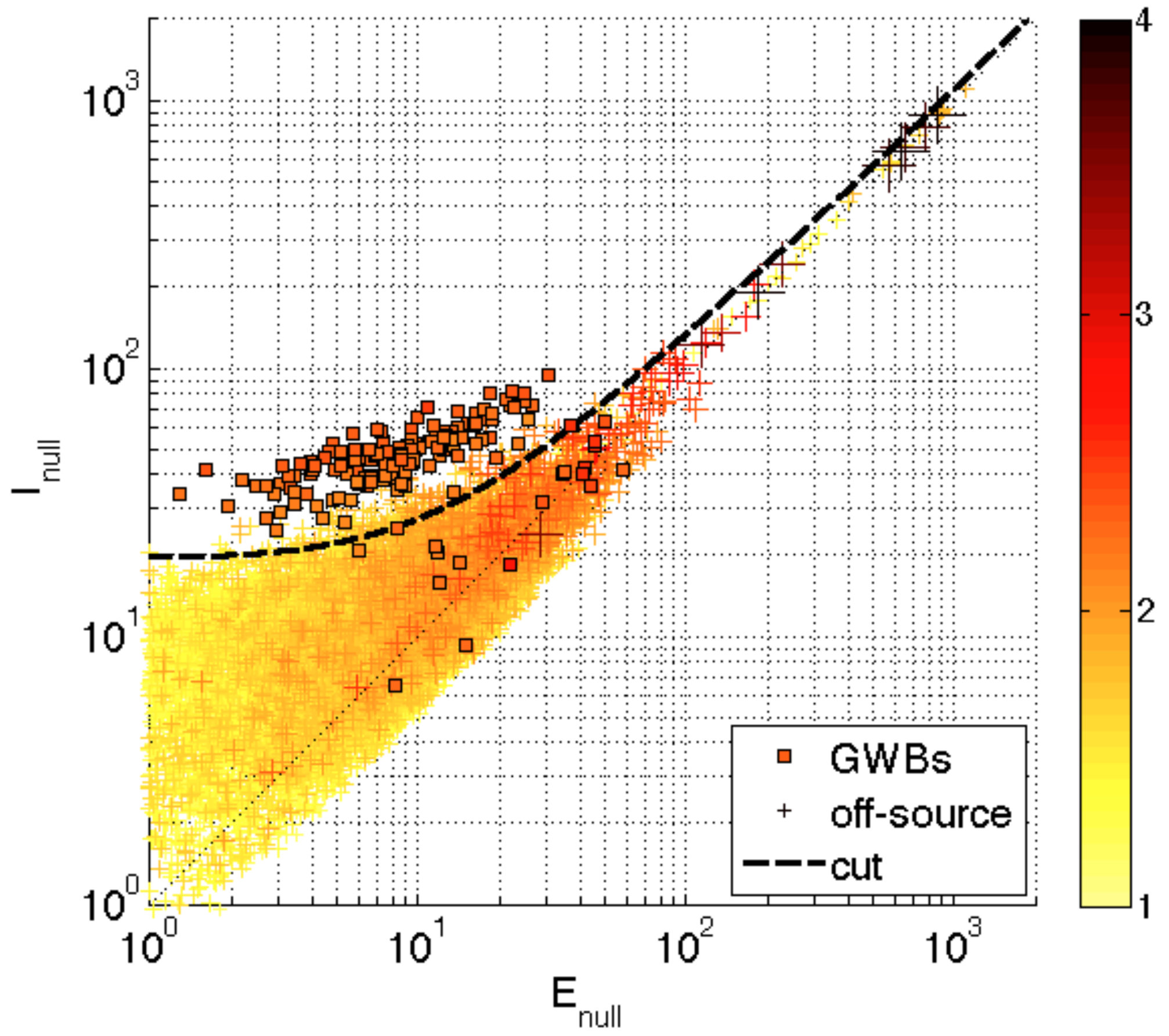}
  \caption{\label{fig:veto} 
$\In$ vs.~$\En$ for clusters produced by background noise ($+$) and 
by simulated gravitational-wave signals ($\Box$).  The color axis is the 
base-10 logarithm of the cluster significance ${\cal S}$.  Loud glitches 
are vetoed by discarding all clusters that fall below the dashed line.
}
\end{center}
\end{figure}

The second type of coherent veto test in \xp~is called the {\em 
median-tracking} veto test.  In this test, the exclusion curve is 
nonlinear and designed to approximately follow the measured 
distribution of background clusters.  

Examination of scatter plots of $I$ vs.~$E$  
for background clusters shows that, while $I \simeq E$ for 
loud glitches, there is a bias to $I > E$ at low amplitudes.  
Furthermore, the width of the distribution of background events 
around the diagonal varies with $E$.  A simple scaling argument 
shows that for large-amplitude uncorrelated glitches we expect 
\be
\langle (E-I)^2 \rangle \propto I \, , 
\ee
Specifically, for a large single-detector glitch $\tilde{g}(f)$, the 
correlation with the noise $\tilde{n}(f)$ in another detector will 
have mean zero and variance $\propto |\tilde{g}|^2 \propto I$.
Consequently, we expect noise events to be scattered about the diagonal 
with a width that is proportional to $I^{1/2}$ (recall that the energies 
are dimensionless quantities).
The median-tracking test uses this information by estimating the median 
value of $I$ as a function of $E$ for background events.  For each cluster 
to be tested, it computes the following simple measure 
$n_\sigma$ of how far 
the cluster is above or below the median: 
\be
n_\sigma \equiv \frac{I - I_\mathrm{med}(E)}{I^{1/2}} \, .
\ee
An event is passed if 
\bea
n_\mathrm{null} & > & r_\mathrm{null} \\
|n_+| & > & r_+ \\
|n_\times| & > & r_\times \, .
\eea
As in the ratio test, the thresholds for each energy 
type are independent and may be specified by the user or selected 
automatically by \xp.

The median function $I_\mathrm{med}(E)$ is estimated as follows.
First, a set of background clusters are binned in $\log_{10}E$ 
and the median values of  $\log_{10}E$ and  $\log_{10}I$ in each bin 
are measured.  A quadratic curve of the form 
\be
\log_{10}I = a (\log_{10}E)^2 + c
\ee
is fit to these sampled medians.  The quadratic is merged smoothly 
to the diagonal $I=E$ above some value of $E$.  This shape is entirely 
ad hoc, but in practice it provides a good fit the observed distribution 
of glitches.

An example of the median-tracking coherent glitch veto is shown in 
Figure~\ref{fig:veto}.  Each plus symbol ($+$) denotes a background 
cluster, colored by its significance $\log_{10}{\cal S}$.  The large 
mass of light points at lower left are weak background noise events.  
The darkly colored points extending along the diagonal to upper 
right are strong background noise events.  Also shown are clusters 
due to a series of simulated gravitational-wave signals added to the 
data, denoted by squares ($\Box$).  Even though many of these 
simulated signals are weaker (lighter) than the strong background noise 
glitches, they are well separated from the background noise 
population in the two-dimensional $(\En,\In)$ space.  
The dashed line shows the coherent veto threshold placed on 
$(\En,\In)$; points below this line are discarded.  Scatterplots 
of $\Ip$ vs.~$\Ep$ and $\Ic$ vs.~$\Ec$ have similar appearance; 
see Section~\ref{sec:grb} for examples.

In addition to the coherent glitch vetoes, clusters may also be rejected 
because they overlap {\em data quality vetoes}.  These are periods 
when one or more detectors showed evidence of being disturbed by 
non-gravitational effects that are known to produce noise glitches.  
Such sources include environmental noise and instabilities in the 
detector control systems.  These data quality vetoes are defined by 
studies of the data independently of \xp, and hence are outside of the 
scope of this report.  See \cite{Blackburn:2008ah,Leroy:2009zz} 
for recent reviews of data quality and detector characterisation efforts 
in the LIGO Scientific Collaboration and the Virgo Collaboration.

\subsection{Triggered search: tuning and upper limits}
\label{sec:triggeredsearch}

We now focus on the strategy for conducting triggered searches with \xp, 
specifically searches for gravitational waves associated with gamma-ray 
bursts (GRBs).  As pointed out by Hayama {\em et al.} \cite{Hayama:2007iz}, 
GRB searches are an excellent case for the application 
of coherent analysis, since the sky position of the source is known 
{\em a priori} to high accuracy.  We can therefore take full advantage of 
coherent combinations of the data streams without the false-alarm 
or computational penalties of scanning over thousands of trial sky directions.

\subsubsection{Detection Procedure}

For the purposes of a search for unmodelled gravitational-wave emission, 
a GRB source is characterised by its sky position $\hat{\Omega}$, the time 
of onset of gamma-ray emission (the {\em trigger time}) $t_0$, and by the 
range of possible time delays $\Delta t$ 
between the gamma-ray emission and the associated gravitational-wave 
emission.  The latter quantity is referred to as the {\em on-source} window 
for the GRB; this is the time interval which is analysed for candidate signals.  
LIGO searches for gravitational wave bursts associated with GRBs 
\cite{Ab_etal:05c,Abbott:2007rh,Ab_etal:08} have traditionally 
used an asymmetric on-source window of $[t_0-120\,\textrm{s},t_0+60\,\textrm{s}]$, which 
is conservative enough to encompass most theoretical models of 
gravitational-wave emission for this source, as well as 
uncertainties associated with $t_0$ \cite{Meszaros:2006rc,Ab_etal:05c}.

In order to claim a detection of a gravitational wave, we need to be able to 
establish with high confidence that a candidate event is statistically 
inconsistent with the noise background.  In \xp~GRB searches, we use the 
{\em loudest event} statistic \cite{Brady:2004gt,Biswas:2007ni} to 
characterise the outcome of the experiment.  
The loudest event is the cluster in the on-source interval that 
has the largest significance (after application of vetoes); let us denote its 
significance by ${\cal S}_\mathrm{max}^\mathrm{on}$.  
%
%
We compare ${\cal S}_\mathrm{max}^\mathrm{on}$ to the cumulative 
distribution $C({\cal S}_\mathrm{max})$ of loudest significances 
measured using background noise (discussed below).
%
We set a threshold on $C({\cal S}_\mathrm{max})$ such that the probability 
of background noise producing a cluster in the on-source interval with 
significance above this threshold is a specific small value (for example, a 1\% 
chance).  The on-source data is then analysed.  If the significance 
$C({\cal S}_\mathrm{max}^\mathrm{on})$ of the loudest cluster is greater than our 
threshold, we consider the cluster as a possible gravitational wave detection.  
We can also set an upper limit on the strength of gravitational-wave 
emission associated with the GRB in question.

In principle, the cumulative distribution $C({\cal S}_\mathrm{max})$ of 
loudest-event significances 
for clusters produced by Gaussian background noise can be estimated 
{\em a priori}.  In practice, however, real detector data is non-Gaussian.  
The most straightforward procedure for estimating the background 
distribution is then simply to analyse additional data from times near the 
GRB, but outside the on-source interval.  These data are referred to as 
{\em off source}. The off-source clusters will not contain a 
gravitational-wave signal associated with the GRB, and so they can be 
treated as samples of the noise background.  In \xp, we divide the 
off-source data into segments of the same length as that used for the 
on-source data, and analyse each segment in exactly the same manner 
as the on-source data (using, for example, the same source direction 
relative to the detectors for computing coherent 
combinations).  For each segment, we determine the significance of the 
loudest event after applying vetoes.  This collection of loudest-event 
significances from the off-source data then serves as the empirical 
measurement of $C({\cal S}_\mathrm{max})$.

In \xp~we typically set the off-source data to be all data within $\pm1.5$ 
hours of the GRB time, excluding the on-source interval.  This time range 
is limited enough so that the detectors should be in a similar state of 
operation as during the GRB on-source interval, but long enough 
to provide typically $\sim$50 off-source segments for sampling 
$C({\cal S}_\mathrm{max})$, thereby allowing estimation of probabilities 
as low as $\sim$2\%.  To get still better estimates of the background 
distribution, we also analyse off-source data after artificially 
time-shifting the data from one or more detectors by different amounts 
ranging from a few seconds to several hundred seconds.  These shifts 
can give up to approximately 1000 times the on-source data for 
background estimation, allowing estimation of probabilities at the 
sub-1\% level.

Networks containing both the LIGO-Hanford detectors, H1 and H2, present 
a special case for background estimation, as local environmental 
disturbances can produce simultaneous background glitches which are 
not accounted for in time slides.  We therefore do not time-shift H1 
relative to H2 unless they are the only detectors operating.  In that  
case, the local probability is computed both with and without time 
slides to allow a consistency check on the background estimation.  
(Triggered searches with second-scale on-source windows have the 
advantage of not requiring time shifts at all; see for example 
\cite{PhysRevLett.101.211102}.) 
In practice, we do not see significant differences due to correlated 
environmental disturbances.  We attribute this robustness to the 
coherent glitch rejection tests described in Section~\ref{sec:veto}.

\subsubsection{Upper Limits}

The comparison of the largest significance measured in the on-source 
data, ${\cal S}_\mathrm{max}^\mathrm{on}$, to the cumulative distribution 
$C({\cal S}_\mathrm{max})$ estimated from the off-source data allows 
us to determine if there is a statistically significant transient 
associated with the GRB.  If no statistically 
significant signal is present, we set a frequentist upper limit on the 
strength of gravitational waves associated with the GRB.  For a given 
gravitational-wave signal model, we define the 90\% confidence level 
upper limit on the signal amplitude as the minimum amplitude for which 
there is a 90\% or greater chance that such a signal, if present in the 
on-source region, would have produced a cluster with significance larger 
than the largest value ${\cal S}_\mathrm{max}^\mathrm{on}$ actually measured.

We adopt the measure of signal amplitude that is standard for LIGO 
burst searches, the root-sum-squared amplitude $h_\mathrm{rss}$, 
defined by 
\begin{eqnarray}
h_\mathrm{rss}
  & = &  \sqrt{\int_{-\infty}^{\infty} \!\! dt \left[ h_+^2(t)+h_\times^2(t) \right]},
         \nonumber \\
  & = &  \sqrt{2 \int_{0}^{\infty} \!\! df \left[ \tilde{h}_+^2(f)+\tilde{h}_\times^2(f) \right]} 
         \, .
\end{eqnarray}
The units of $h_{rss}$ are Hz$^{-1/2}$, the same as for amplitude spectra, 
making it a convenient quantity for comparing to detector noise curves.   
For narrow-band signals, the $h_\mathrm{rss}$ can also be linked to the 
energy emitted in gravitational waves under the assumption of isotropic 
radiation via \cite{Ri:04}
\begin{equation}\label{Eiso}
 E^\mathrm{iso}_\mathrm{GW} \simeq \frac{\pi^2 c^3}{G}D^2 f_0^2 h_\mathrm{rss}^2 \, ,
\end{equation}
where $D$ is the distance to the source and $f_0$ is the dominant 
frequency of the radiation. One drawback of $h_\mathrm{rss}$ is that it 
does not involve the detector sensitivity (either antenna response or 
noise spectrum). As a result, upper limits phrased in terms of 
$h_\mathrm{rss}$ will depend on the family and frequency of waveforms 
used, and also on the sky position of the source. 

To set the upper limit, we need to determine how strong a real 
gravitational-wave signal needs to be in order to appear with a given 
significance.  We do this using a third set of clusters, one which contains 
sample gravitational-wave signals.  Specifically, we repeatedly 
re-analyse the on-source data after adding (``injecting'') simulated 
gravitational-wave signals to the data from each detector.  The data 
is then analysed as before, producing lists of clusters.  The significance 
associated with a given injection is the largest significance of all clusters 
that were observed within a short time window (typically 0.1s) of the 
injection time, after applying vetoes.

The procedure for setting an upper limit is:
\begin{enumerate}
\item
Select one or more families of waveforms for which the upper limit 
will be set.  For example, a common choice in LIGO is linearly 
polarized, Gaussian-modulated sinusoids (``sine-Gaussians'') with 
fixed central frequency and quality factor, and random peak time 
and polarization angle.
\item
Find the significance ${\cal S}_\mathrm{max}^\mathrm{on}$ of the loudest event 
in the on-source data, after applying the coherent glitch veto 
(Section~\ref{sec:veto}) and any data-quality vetoes.
\item \label{item:3}
For each waveform family:
\begin{enumerate}
\item \label{item:3a}
Generate random parameter values for a large number of waveforms 
from the family ({\em e.g.}, specific peak times and polarization angles 
for the sine-Gaussian case), and with fixed $h_\mathrm{rss}$ amplitude.  
\item \label{item:3b}
Add the waveforms one-by-one to the on-source data, and determine 
the largest significance of any surviving cluster (after vetoes) associated 
with each injection.  
\item \label{item:3c}
Compute the percentage of the injections that have 
${\cal S} \ge {\cal S}_\mathrm{max}^\mathrm{on}$.
\item
Repeat \ref{item:3a}--\ref{item:3c} using the same waveform family 
but with different $h_\mathrm{rss}$ amplitudes.  The 90\% 
confidence-level upper limit is that $h_\mathrm{rss}$  
for which 90\% of the injections have ${\cal S} \ge {\cal S}_\mathrm{max}^\mathrm{on}$.
\end{enumerate}
\end{enumerate}

\subsubsection{Tuning and Closed-Box Analyses} 
\label{sec:tuning}

The sensitivity of the pipeline is determined by the relative significance 
of the clusters produced by real gravitational-wave signals to those 
produced by background noise.  This in turn depends on the details of 
how the analysis is carried out.  In particular, the thresholds used for 
the coherent glitch rejection tests will have a significant impact on the 
sensitivity.  Too low a threshold will allow background noise glitches to 
survive, and possibly appear louder than a real gravitational-wave signal.  
Too high a threshold may reject the gravitational-wave signals we seek.

To improve the sensitivity of \xp~searches, we tune the coherent glitch 
test to optimize the trade-off between glitch rejection and signal 
acceptance.  We do this using a {\em closed-box} analysis.  A closed-box 
analysis estimates the pipeline sensitivity using the off-source and injection 
data, but not the on-source data.  This {\em blind} tuning avoids the 
possibility of biasing the upper limit.

The procedure used for a closed-box analysis follows that used for 
computing an upper limit, except that an off-source segment is used as 
a substitute for the true on-source segment.  We then test different 
thresholds for the coherent veto tests, and select the threshold set 
that gives us the best average ``upper limit'' estimated from the 
off-source segments.  Specifically, we do the following:
\begin{enumerate}
\item
For each coherent veto test ($\Ep$ vs.~$\Ip$, $\Ec$ vs.~$\Ic$, 
$\En$ vs.~$\In$) we select a discrete set of trial veto thresholds to test.
\item
The off-source segments and the injection clusters are divided randomly 
into two equal sets: one for tuning, and one for upper-limit estimation.
\item\label{titem:3}
For each distinct combination of trial thresholds 
$(r_+,r_\times,r_\mathrm{null})$, we do the following:
\begin{enumerate}
\item\label{titem:3a}
We apply the coherent veto test (and 
any data quality vetoes) to the background clusters from each of the 
tuning off-source segments.  The collection of loudest surviving events from 
each segment gives us $C({\cal S}_\mathrm{max})$ for that set of 
trial thresholds.
\item
We determine the off-source segment that gives the loudest event 
closest to the $95^\mathrm{th}$ percentile of the off-source 
${\cal S}_\mathrm{max}$ (i.e., closest to 
$C({\cal S}_\mathrm{max})=0.95$).  This off-source 
segment is termed the {\em dummy on-source segment}.  
(Different background segments may serve as the dummy 
on-source for different trial values of the coherent veto 
thresholds.)
\item\label{titem:3c}
The dummy on-source clusters and the tuning injection clusters are read, 
and the coherent vetoes and data-quality vetoes are applied to each.
The upper limit is computed, treating the dummy clusters as the true 
on-source clusters.
\end{enumerate}
\item
The final, tuned veto thresholds are 
the ones that give the lowest upper limit based on the dummy on-source 
clusters.  (If testing multiple waveform families, the upper limits 
may be averaged across families for deciding the optimal tuning.)
\item\label{item:5}
To get an unbiased estimate of the expected upper limit, we apply 
the tuned vetoes to the second set of off-source and injection clusters, 
that were not used for tuning.
Steps \ref{titem:3a} -- \ref{titem:3c} are repeated using the final 
thresholds, and using the $50^\mathrm{th}$ percentile of 
${\cal S}_\mathrm{max}$ to choose the dummy on-source segment.  
The upper limit estimated from the dummy on-source segment in this 
second data set is the predicted upper limit for the GRB; equivalently, 
it may be interpreted as the sensitivity of the search.
\end{enumerate}
We choose the $95^\mathrm{th}$ percentile of ${\cal S}_\mathrm{max}$ 
for tuning to focus on eliminating the tail of high-significance 
background glitches.  This is a deliberate choice, since to be 
accepted as a detection, a GWB will need to stand well clear of the 
background.
We choose the $50^\mathrm{th}$ percentile of ${\cal S}_\mathrm{max}$ 
as the dummy on-source value for sensitivity estimation because this 
is our best prediction for the typical value of ${\cal S}_\mathrm{max}$ 
in the on-source data under the null hypothesis.  
Separate data sets are used in tuning and sensitivity estimation to 
avoid bias from tuning the cuts on the same data used to estimate the 
sensitivity.    
The data set used for closed-box sensitivity estimates is later re-used 
for computing event probabilities and upper limits for the ``open-box'' 
(true on-source) data; this introduces no bias because no tuning decisions 
are made based on the closed-box sensitivity estimate. 


In \xp, the tuning and upper limit calculations are automated.  
The closed-box analysis is performed first using a pre-selected range of 
trial thresholds for the coherent glitch test.  
A web page is generated automatically reporting the details of the 
closed box analysis, including the optimized threshold values and 
the predicted upper limits.  
For the S5/VSR1 search, the user re-runs the post-processing 
on the on-source data with the fixed optimized thresholds, and  
another web page report is generated listing detection candidates 
and upper limits.
For the S6/VSR2 search, we propose to automate this ``box opening'' 
as well, so that the on-source events are scanned for candidate GWBs 
immediately once the closed-box tuning analysis has finished.

\subsubsection{Statistical and Systematic Errors}
\label{sec:errors}

There are several sources of error that can affect our analysis.  
The principal ones are calibration uncertainties (amplitude and 
phase response of the detectors, and relative timing errors), and 
uncertainty in the sky position of the GRB. 

\xp~is able to account for these effects automatically in tuning and 
upper limit estimation.
Specifically, \xp's built-in simulation engine for injecting GWB signals 
is able to perturb the amplitude, phase, and time delays for each injection 
in each detector.  The perturbations are drawn from Gaussian distributions 
with mean and variance matching the calibration uncertainties for each 
detector.  Furthermore, the GRB sky position can be perturbed in a random 
direction by a Gaussian-distributed angle with standard deviation set to 
the GRB error box width reported by the GCN.  
Tuning and upper limits based on the perturbed injections are effectively 
marginalized over these sources of error.

For the \runone~GRB search, the capability for perturbed injections 
was not available at the time of the original data analysis, and so 
the impact of the errors was estimated by re-analysis of a small 
subset of the full GRB sample.  For the \runtwo~search, we  
include calibration and sky-position uncertainties in simulations 
for all GRBs from the beginning, removing the need to do any additional  
error analysis.

\section{GRB 031108}
\label{sec:grb}

GRB 031108 \cite{GRB031108} was a long GRB observed by Ulysses, 
Konus-Wind, Mars Odyssey-HEND and GRS, and RHESSI.  As observed by 
Ulysses, it had a duration of approximately 22 seconds, a 25-100 keV 
fluence of approximately $2.5\times10^{-5}$ erg/cm$^2$, and a peak 
flux of approximately  $1.8\times10^{-6}$ erg/cm$^2$ s over 0.50 seconds.  
It was triangulated to a 3-sigma error box with approximate area 
1600 square arcminutes with center coordinates 4h 26m 54.86s, 
$-5^\circ \, 55^\prime \, 49.00^{\prime\prime}$. 

GRB 031108 occurred during the third science run of the LIGO Scientific 
Collaboration (``S3'').  At that time, the two LIGO-Hanford detectors 
H1 and H2 were operating, while the Livingston detector L1 was not.  
A search for gravitational waves associated with the GRB was performed 
using a cross-correlation algorithm, and reported in Abbott {\em et al.} 
\cite{Ab_etal:08}.

To demonstrate \xp, we perform a closed-box analysis \footnote{
We restrict ourselves to closed-box results because the policies governing 
LIGO data use do not permit the publication of open-box analysis results 
in methodological papers.
} 
of the LIGO 
H1-H2 data to search for gravitational waves associated with GRB 
031108.  We tune the search and estimate its sensitivity to 
gravitational-wave emission as discussed in Section~\ref{sec:triggeredsearch}, 
using the same simulated waveforms as in Abbott {\em et al.}.  
We compare the sensitivity results to those of the cross-correlation 
search in Abbott {\em et al.}  We estimate the 90\% confidence upper 
limits from \xp~to be typically 40\% lower than those from the 
cross-correlation search.

\subsection{Analysis}
\label{sec:network}

\begin{figure}[htb]
\begin{center}
  \includegraphics[width=3in]{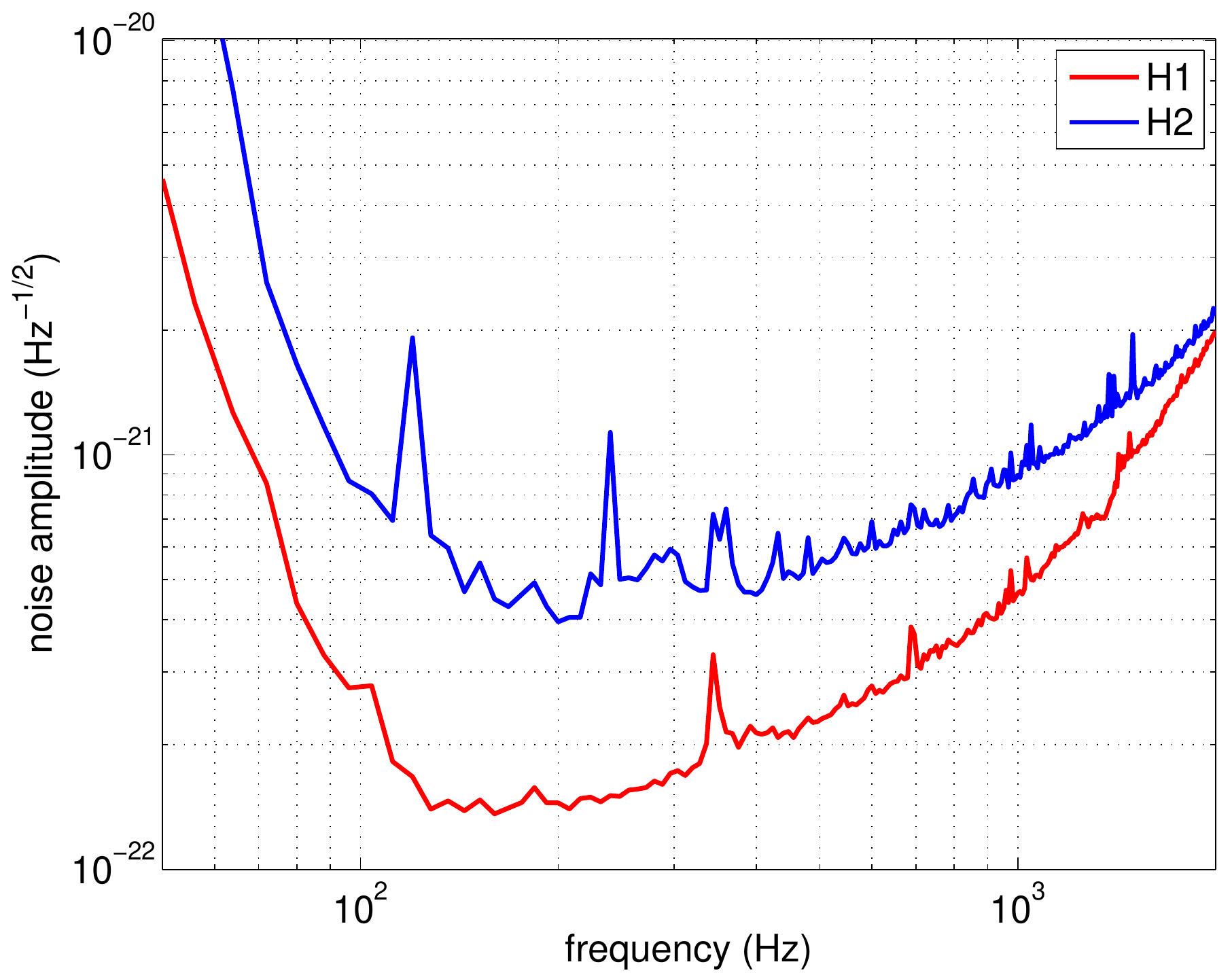}
  \caption{\label{fig:spectra} 
  Noise spectra of the H1 and H2 detectors at the time of 
  GRB 031108 as estimated by \xp.
}
\end{center}
\end{figure}

At the time of GRB 031108, the two LIGO Hanford detector H1 and H2 
were operating.  Figure~\ref{fig:spectra} shows the noise level in 
the detectors at that time.

\begin{figure}[htb]
\begin{center}
  \includegraphics[width=3in]{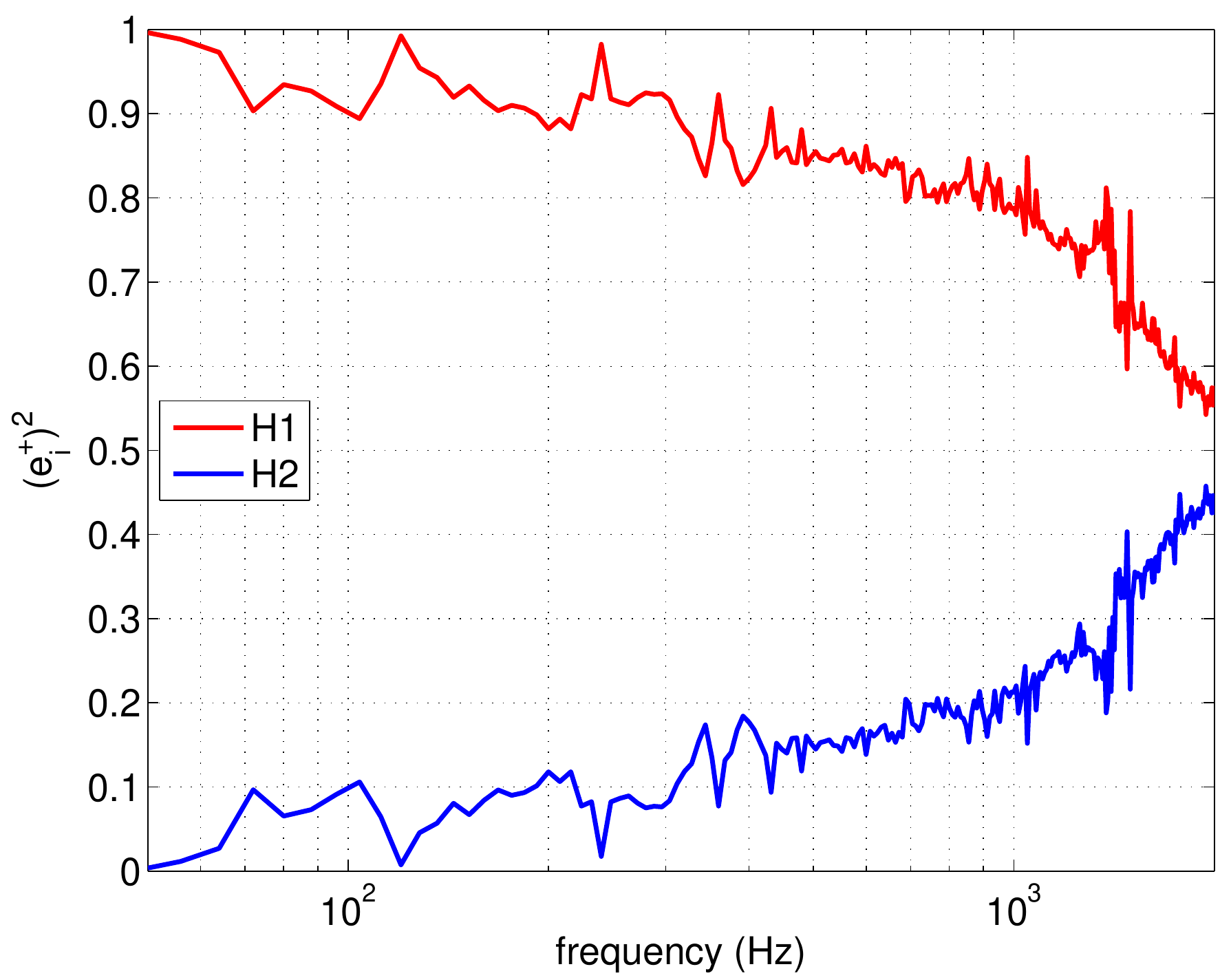}
  \caption{\label{fig:cohcomb} 
  Normalized contributions  
  to $\FplusHat\equiv\FplusDP/|\FplusDP|$ (in the DPF).  Both detectors have the same antenna 
  response, so the coherent weighting at each frequency is determined 
  entirely by the relative noise spectra.  Because $\FcrossDP=0$ for 
  this network, the normalized contribution of detector $i$ to the null stream is 
  simply $1-(e_i^+)^2$; i.e., identical to this figure with the 
  H1 and H2 curves swapped.
}
\end{center}
\end{figure}

Since the H1 and H2 detectors have identical antenna responses, 
the network is sensitive to only one of the two gravitational-wave 
polarizations from any given sky direction.  In the DPF, 
this means that $\FcrossDP=0$.  As a consequence, the cross energy 
also vanishes identically, $\Ec=0$, and $\Esl=\Ep$.  Each event 
cluster is therefore characterised by the two coherent energies 
$\Ep$ and $\En$, and their associated incoherent components $\Ip$ 
and $\In$.  Figure~\ref{fig:cohcomb} shows the weighting factors 
$\FplusHat$ as a function of frequency.

\begin{figure}
\begin{center}
  \includegraphics[width=3in]{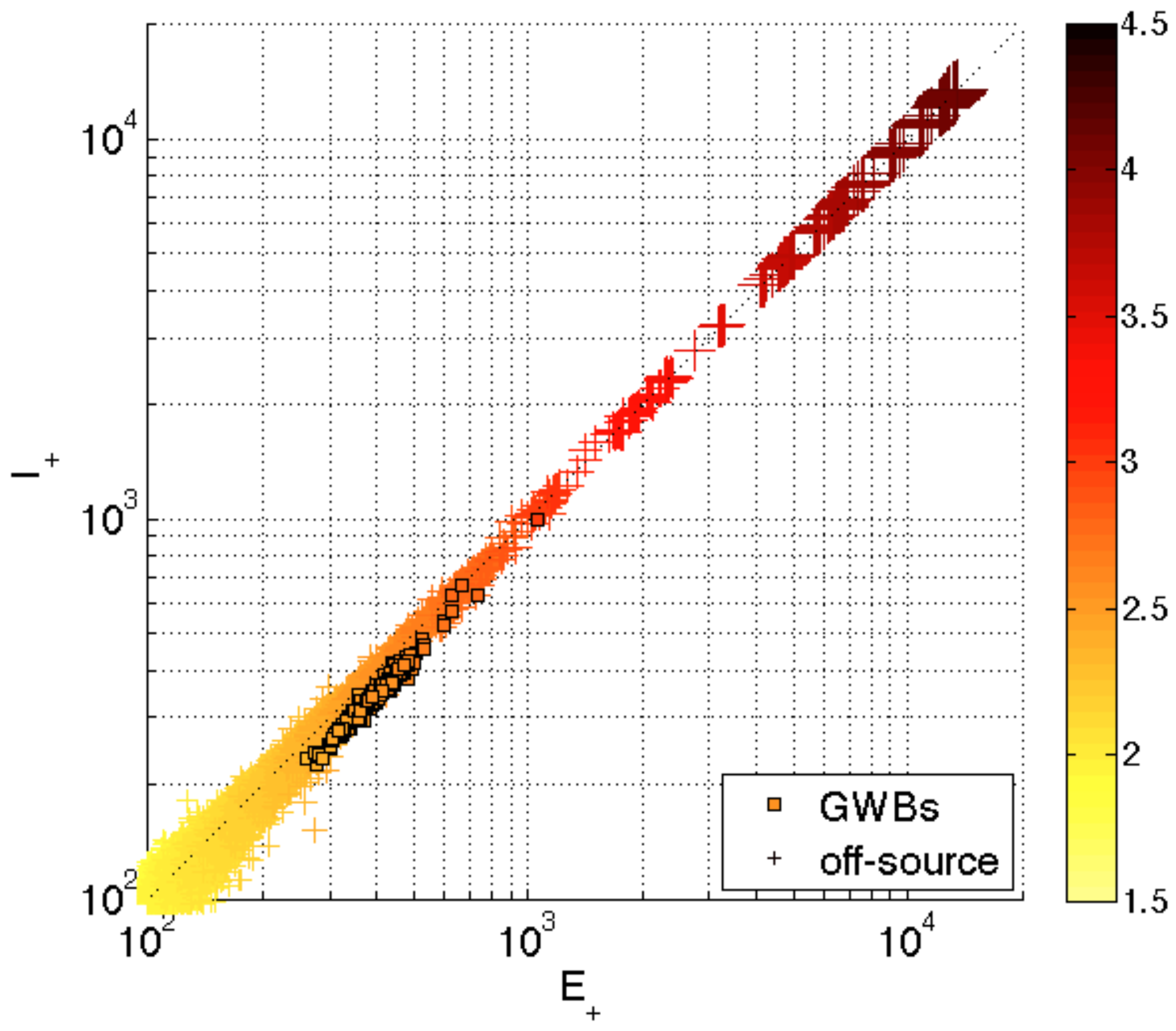}
  \includegraphics[width=3in]{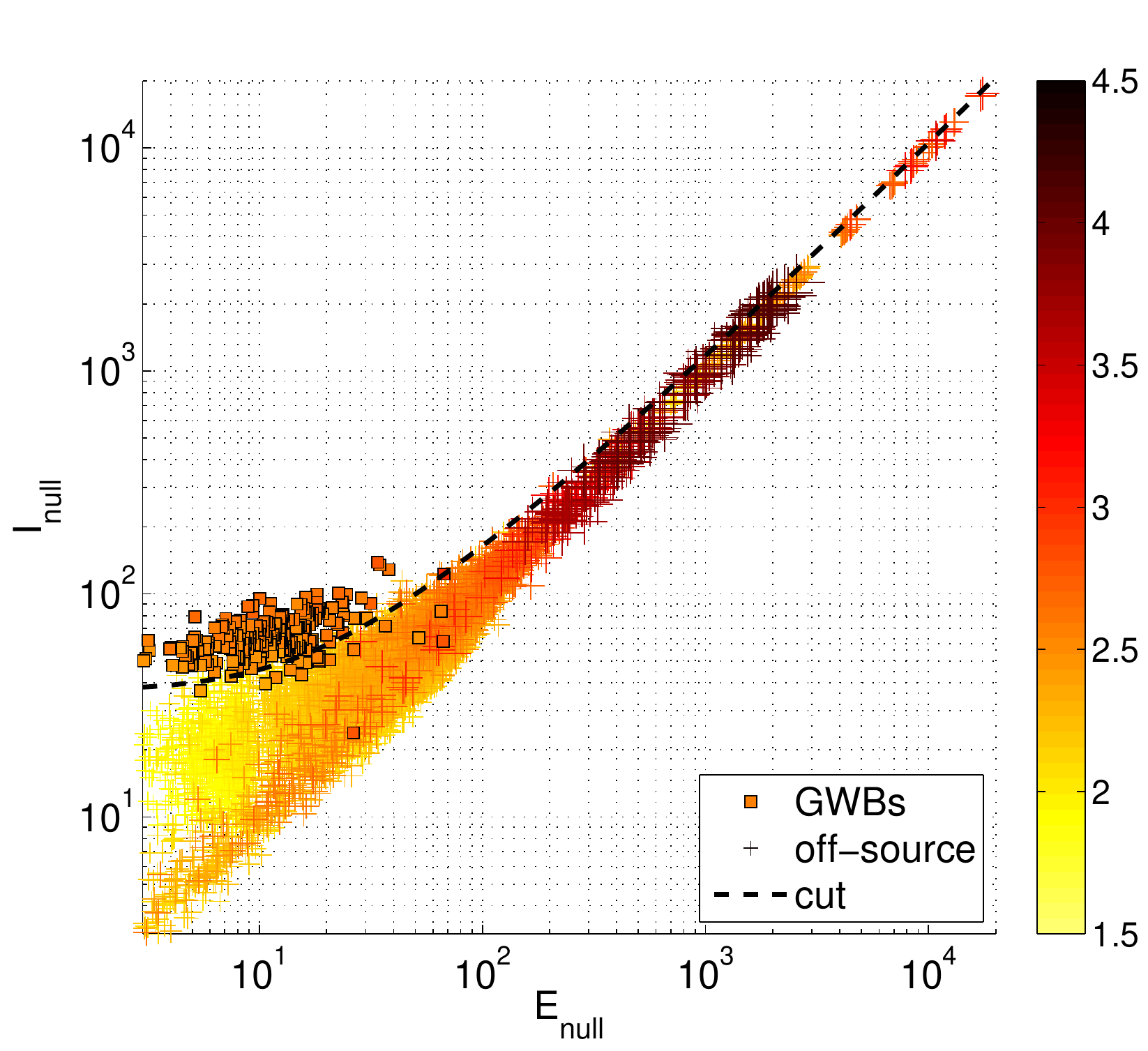}
  \caption{\label{fig:offscatter} 
  Scatter plots of off-source ($+$) and simulation ($\Box$) cluster 
  likelihoods: $\Ip$ vs.~$\Ep$ (top) and $\In$ vs.~$\En$ (bottom).  
  The color denotes $\log_{10}({\cal S})$.
  Loud background triggers fall close to the diagonal.
  Simulated gravitational waves also fall close to the diagonal for 
  $\Ip$ vs.~$\Ep$, but above the diagonal for $\In$ vs.~$\En$. 
  The dashed line denotes the coherent consistency threshold on 
  $(\En,\In)$ that is selected by \xp's automated tuning procedure; 
  points below this line are discarded.  This test rejects the 
  majority of the loud off-source clusters, while accepting most of 
  the simulated gravitational wave clusters, even if the GWB 
  significiance is typical of background events.  The simulated signals 
  in this plot have $h_\mathrm{rss} = 6.3\times10^{-21}\mathrm{Hz}^{-1/2}$, 
  approximately equal to the upper limit estimated from the closed-box 
  analysis.  
}
\end{center}
\end{figure}

\xp~was run on all data within $\pm1$hr of the GRB time for 
background estimation.  Clusters were generated using Fourier 
transform lengths of 1/8s, 1/16s, 1/32s, 1/64s, 1/128s, and 1/256s.  
Figure~\ref{fig:offscatter} shows scatter plots of $\Ip$ vs.~$\Ep$ 
and $\In$ vs.~$\En$ for the half of the off-source clusters that 
were used for upper limit estimation ({\em i.e.}, after tuning).  Also shown are 
the clusters produced by simulated sine-Gaussian GWBs at 150 Hz, 
one of the types tested in \cite{Ab_etal:08}.  These injections had 
amplitudes of $6.3\times10^{-21}\mathrm{Hz}^{-1/2}$, approximately 
equal to the $h_\mathrm{rss}$ upper limit estimated from the 
closed-box analysis.

\begin{figure}[htb]
\begin{center}
  \includegraphics[width=3in]{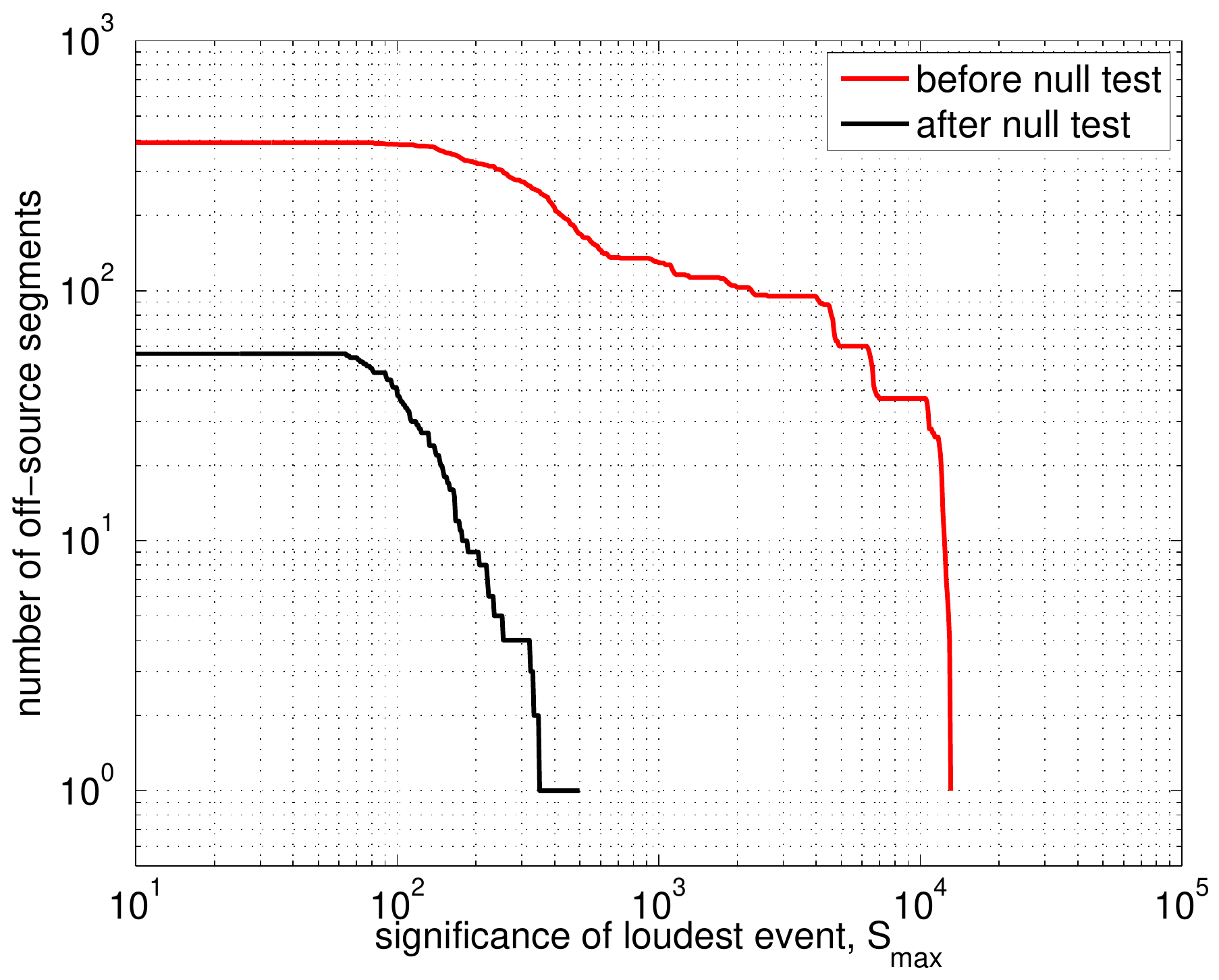}
  \caption{\label{fig:bckgrd} 
  Distribution of the loudest event significance 
  ${\cal S}_\mathrm{max}$ seen in each of the off-source 
  segments used for upper limit estimation, before and after the 
  coherent glitch rejection test.  Only 56 of the 391 off-source 
  segments have events that survive the test.
}
\end{center}
\end{figure}

As expected, loud background triggers fall close to the diagonal 
in both of these plots. The simulated gravitational waves also fall 
close to the diagonal for $\Ip$ vs.~$\Ep$; this is due to the fact 
that H2 is significantly less sensitive than H1 and so receives very 
little weighting in the calculation of $\Ep$.  In turn, this means 
that the H1-H2 cross terms in $\Ep$ are small compared to the H1-H1 
term, so that $\Ep$ is dominated by the diagonal components and so 
is very similar to $\Ip$.  For the null stream, however, the 
weightings are reversed, and H2 is weighted higher than H1.  As a 
consequence, gravitational waves lie above the diagonal in the 
$\In$ vs.~$\En$ plot, and it is possible to separate the injections 
from the background clusters in $(\En,\In)$ space.  \xp's automated 
tuning procedure recognizes both of these facts; when run using the 
median-tracking veto test, it estimates that the best sensitivity  
will come from requiring a threshold of $r_+=5$ on $(\En,\In)$, and 
imposing no condition on $\Ip$ vs.~$\Ep$.  The $(\En,\In)$ threshold 
is indicated in Figure~\ref{fig:offscatter} by the dashed line; 
points below this line are discarded.  As can be seen, this test 
rejects the majority of the loud off-source clusters, while 
accepting most of the simulated gravitational wave clusters.  The 
off-source clusters that survive the test tend to be of low 
significance, and therefore will not affect the loudest-event upper 
limit.  Figure~\ref{fig:bckgrd} shows the distribution of 
${\cal S}_\mathrm{max}$ before and after the null-stream test.

\begin{figure}[b]
\begin{center}
  \includegraphics[width=3in]{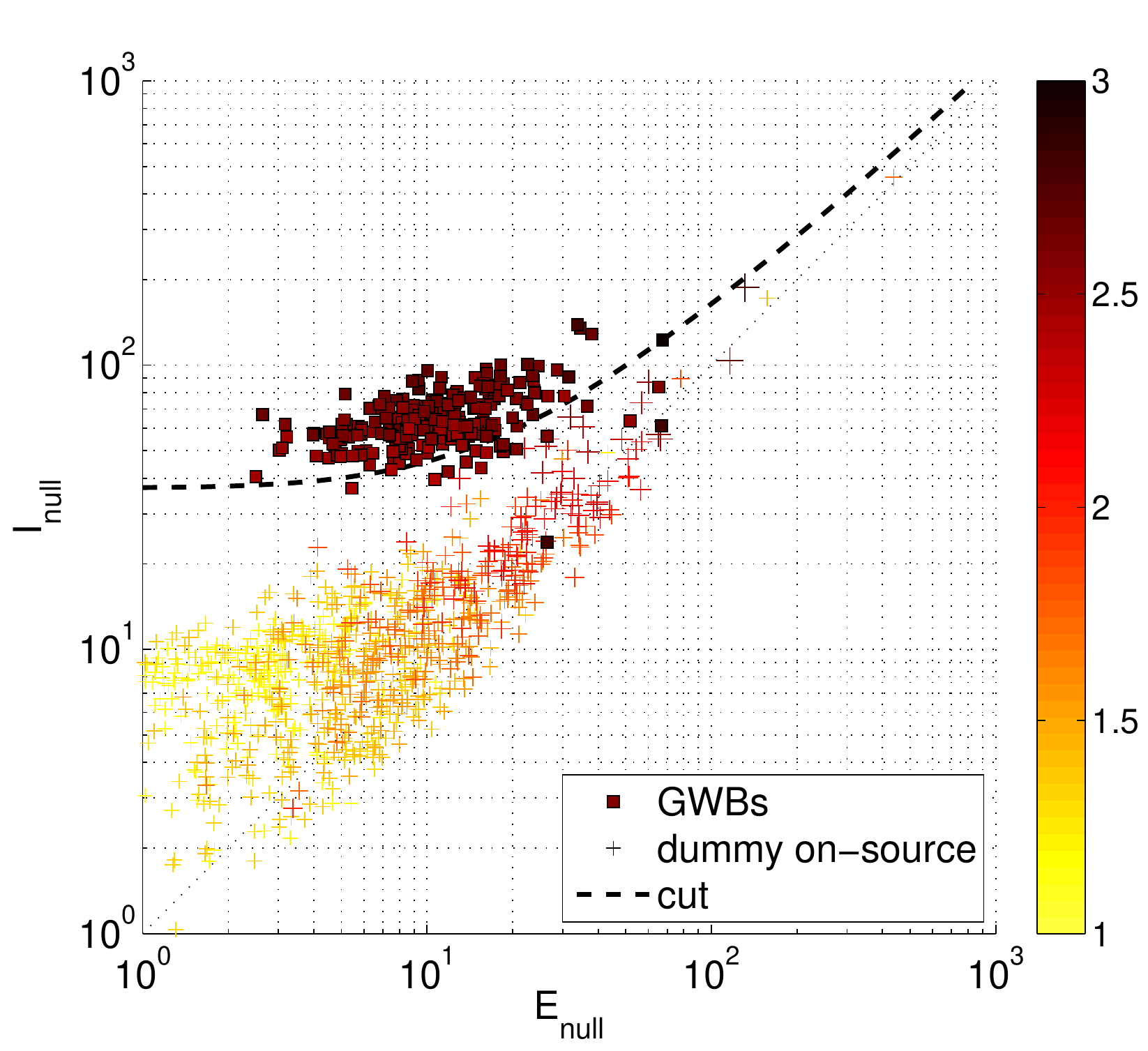}
  \caption{\label{fig:onscatter} 
  $\In$ vs.~$\En$ scatter plot of the dummy on-source ($+$) and 
  simulation ($\Box$) cluster likelihoods used to estimate the 
  upper limit.  The color denotes $\log_{10}({\cal S})$.
  No background events survive the coherent consistency test.
  The simulated signals in this plot have 
  $h_\mathrm{rss} = 6.3\times10^{-21}\mathrm{Hz}^{-1/2}$, 
  approximately equal to the estimated upper limit.
}
\end{center}
\end{figure}

The closed-box analysis discussed in Section~\ref{sec:triggeredsearch} was used 
to tune the coherent veto test and estimate the expected upper limit 
from \xp.  Figure~\ref{fig:onscatter} shows a scatter plot of the ``dummy'' 
on-source clusters.  Recall that the dummy on-source region is selected 
as the background segment that gives the median loudest event surviving 
the coherent veto test.  It therefore represents the expected typical result 
under the null hypothesis, averaging over noise instantiations, and so is a 
more robust way to estimate the pipeline sensitivity than, {\em e.g.}, picking 
a random segment (or even the on-source segment).

\begin{figure}[b]
\begin{center}
  \includegraphics[width=3in]{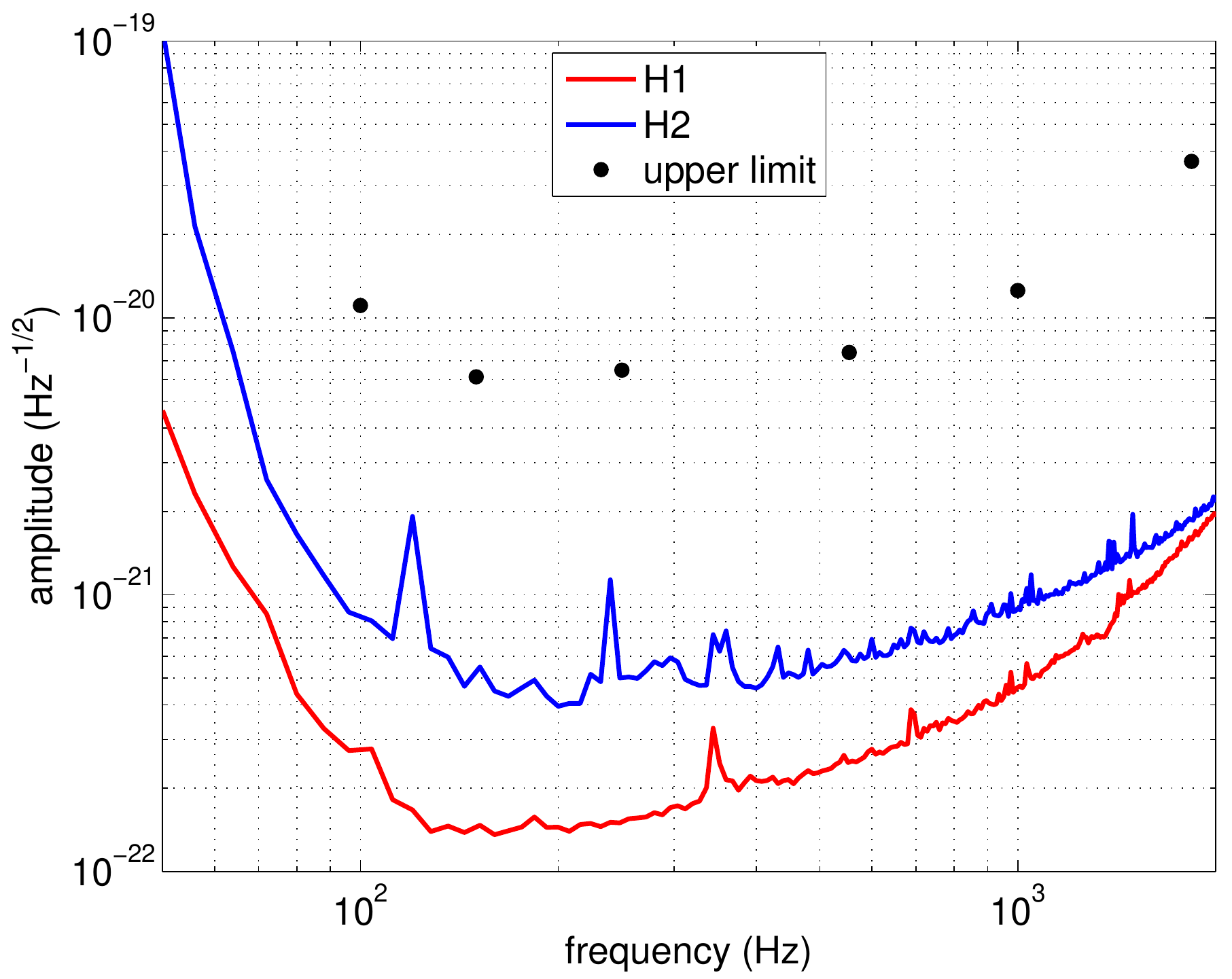}
  \caption{\label{fig:uls} 
  90\%-confidence level upper limits on the GW amplitude ($\bullet$) 
  from \xp~for narrow-band circularly polarized sine-Gaussian bursts.  
  The detector noise spectra are also shown for reference.  
}
\end{center}
\end{figure}

The predicted $h_\mathrm{rss}$ upper limits at 90\%-confidence for 
narrow-band sine-Gaussian waveforms of different central frequencies 
are shown in Table~\ref{tab:ULs} and Figure~\ref{fig:uls}.
Table~\ref{tab:ULs} also shows the actual upper limits from the 
cross-correlation search reported 
in \cite{Ab_etal:08}.  The predicted \xp~sensitivity is approximately 
a factor of 1.7 better than that of the cross-correlation pipeline, 
corresponding to an increase in search volume of a factor of $1.7^3\simeq5$.
Similar improvements were seen in the open-box analysis of GRBs in 
the \runone~run (2005-2007) \cite{P0900023}.

\begin{table*}
\begin{center}
\begin{tabular}{|l|r|r|r|r|r|r|}
\hline
  frequency (Hz)  &  100   &     150    &    250     &   554     &  1000     &  1850    \\
\hline
  cross-correlation  &  18.4  &     11.3  &     10.9   &    12.5    &   20.4     &  51.5  \\
  \xp                &  11.1  &      6.1  &      6.5   &     7.5    &   12.6     &  36.7 \\
\hline
\end{tabular}
\caption{\label{tab:ULs} 
Estimated $h_\mathrm{rss}^{90\%}$ amplitude upper limits from 
\xp~and the best upper limits from the actual cross-correlation search 
\cite{Ab_etal:08}.  The units are $10^{-21} \mathrm{Hz}^{-1/2}$.
The simulated waveforms are circularly polarized sine-Gaussians 
as described in \cite{Ab_etal:08}.
}
\end{center}
\end{table*}

As can been seen in Figure~\ref{fig:uls}, the limiting amplitudes 
for this GRB track the noise spectrum of H2, and correspond to a 
matched-filter signal-to-noise ratio of approximately 5 in H2. 
This occurs because the sensitivity of the analysis is limited by the 
coherent glitch rejection test.  This test requires a measurable 
correlation between the detectors, which in turn requires that the 
GWB have some minimal signal-to-noise ratio in each.  This behaviour is typical of 
tuning using the $95^\mathrm{th}$ percentile of ${\cal S}_\mathrm{max}$, which 
is an aggressive choice designed to suppress the loud background.  
While the upper limits tend to be limited by such strong background 
rejection, our ability to detect a GWB is enhanced, since a GWB candidate 
will undoubtedly need a significance higher than some very high percentile 
of the background to be claimed as an actual gravitational wave.

The factor of 1.7 sensitivity improvement of \xp~relative to the 
cross-correlation search in \cite{Ab_etal:08} can be attributed 
in part to two factors.  We estimate that a factor of approximately 
1.3 comes from using $\Esl$ rather than the cross-correlation as 
the detection statistic.  $\Esl$ includes the auto-correlation terms 
($\fdata^{\vphantom{*}}_\mathrm{H1}\fdata^*_\mathrm{H1}$, 
$\fdata^{\vphantom{*}}_\mathrm{H2}\fdata^*_\mathrm{H2}$) 
in addition to the cross-correlation terms 
($\fdata^{\vphantom{*}}_\mathrm{H1}\fdata^*_\mathrm{H2}$) 
when combining the H1 and H2 data streams.  
This gives a net increase in the signal-to-noise ratio.
More precisely, one can compute the ratio of the expected contribution 
to $\Esl$ due a GWB to the standard deviation in $\Esl$ due to Gaussian 
noise; see Section~\ref{sec:statistics}.  Performing the same calculation 
for the cross-correlation statistic, one finds the per-pixel ratio 
for $\Esl$ to be $1.8\sim1.3^2$ times larger than that for the 
cross-correlation (assuming a 2:1 ratio in the noise amplitudes for H2:H1).
Another factor of $\sim1.2$ can be attributed to the 
clustering, which restricts the likelihood calculation to pixels that 
show significant signal power (and thus tending to exclude pixels that 
contain only background noise).  The cross-correlation statistic 
in \cite{Ab_etal:08} was computed on a minimum time-frequency volume 
(number of pixels) of approximately 50.  By contrast, the typical cluster 
size in \xp~was found to be 10-30 for injections at the 90\% upper limit 
amplitude.  As seen in Section~\ref{sec:statistics} and \cite{AnBrCrFl:01}, 
the amplitude sensitivity in Gaussian noise scales as $N^{-1/4}$.  The  
factor of $\sim2$ smaller number of pixels used by \xp~should therefore 
give a factor of $\sim2^{1/4}=1.2$ sensitivity improvement.  Combined 
with the previous factor of 1.3 gives a total improvement of about 1.6.  
While this is very close to the average measured improvement, one 
should keep in mind that these rough estimates have not properly accounted 
for the non-Gaussianity of the background (which will decrease the 
sensitivity of both pipelines), or for the tendency of the 
coherent glitch rejection test to limit the \xp~sensitivity in the absence 
of strong background glitches.  These other effects are presumably also important.

\section{Autonomous Running}
\label{sec:auto}

\xp~has been used to process data from \runone~(2005-2007).  
This is an ``offline'' search, being completed almost two years after 
the last of the GRBs in question was observed.  In parallel, \xp~is 
being improved for the \runtwo~run, which started in July 2009.  
Our goal for \runtwo~is fully autonomous running, with a complete analysis 
of each GRB within 24 hours of the trigger.  To achieve this goal
requires automatic triggering of \xp.

\subsection{Automated launch of \xp~by GCN triggers}

Most of the information for sources which are analyzed by various 
externally triggered burst searches in LIGO-Virgo come from the 
GRB Coordinates Network (GCN) \cite{gcn}.  GCN notices and circulars 
are received in real time by LIGO-Virgo, and the information needed 
for the search analyses are parsed automatically by perl scripts 
which are launched each time a GCN notice or 
circular is received.  The information parsed includes: the time 
and date of the event, the source position (right ascension and 
declination), the position error, and the duration of the event.  
For each source, these parameters are compiled and written 
to a trigger file.

Concurrently, a perl script runs at a central computing site 
and regularly checks if there are new source 
events listed in the trigger file.  When there are new triggers, 
the script checks for availability of the LIGO-Virgo data which 
are necessary for analyzing the source.  If the needed data are 
available, the script launches \xp~event-generation jobs 
(which include simulation and off-source analyses) on the computing 
cluster.  These jobs are monitored continuously to automatically 
determine when the jobs have finished.  Once they are completed, 
the post-processing (tuning and detection/upper limit) 
jobs are automatically launched and likewise monitored.  
Successful completion of these steps results 
in a web page in which the results of the analysis are presented, 
and an email notification being sent to human analysts. 
Additionally, the scripts which monitor the status of the search 
and post-processing jobs log that progress 
for each source event and regularly write this 
information to a summary status web page.
These GCN parsing and triggering scripts are now 
operational, and \xp~is currently autonomously analysing 
GRBs from the {\em Swift} \cite{swift04} satellite.  
Open-box results are available in as little as 6 hours following 
a GCN alert.  

Other modifications currently being made to \xp~focus on the 
larger sky position error boxes from the Fermi satellite \cite{fermi}.  
For \runtwo, most of the GRB triggers come from the GBM instrument on 
Fermi, which gives a typical position uncertainty of several degrees.  
This is much larger than the typical uncertainty of a few arcmin 
for GRBs from {\em Swift} in \runone.
The \xp~launch scripts are currently being modified to set up a grid 
of sky positions covering this error region, and the handling of 
events is being modified to minimize the additional computational 
time required.  
Finally, the suite of simulated waveforms has been expanded to include 
binary neutron star and black--hole--neutron--star binary inspirals, 
since these systems are widely thought to be the progenitors of short GRBs.

\section{Summary}
\label{sec:summary}

\xp~is a software package designed to perform autonomous searches 
for gravitational-wave bursts associated with astrophysical triggers 
such as gamma-ray bursts.
It performs a fully coherent analysis of data from arbitrary 
networks of detectors to sensitively search small patches of the sky 
for gravitational-wave bursts. 
\xp~features automated tuning 
of background rejection tests, and a built-in simulation engine with 
the ability to simulate effects such as calibration 
uncertainties and sky position errors.
\xp~can be launched automatically by receipt of a GCN email, performing 
a complete analysis of data, including tuning and identification of GWB 
candidates, without human intervention.
Each astrophysical trigger is analysed as a separate search, with 
background estimation and tuning performed using independent data 
samples local to the trigger. 
In a test on actual detector data for a real GRB, we find that \xp~is 
sensitive to signals approximately a factor of 1.7 weaker than those 
detectable by the cross-correlation technique used in previous LIGO 
searches.  
\xp~has recently been used for the analysis of GRBs from from the 
LIGO-Virgo \runone~run, and is currently running autonomously 
during the \runtwo~run to search for gravitational waves associated 
GRBs observed electromagnetically.  Our goal is the rapid identification 
of possible GWBs on time scales short enough to prompt additional 
follow-up observations by other observatories.

\section*{Acknowledgements}

We thank Kipp Cannon and Ray Frey for valuable comments on an 
earlier draft of this paper. 
PJS and GJ were supported in part by STFC grant PP/F001096/1.  
For MT, the research was performed at the Jet Propulsion Laboratory, 
California Institute of Technology, under contract with the National 
Aeronautics and Space Administration.  MT was supported under research 
task 05-BEFS05-0014.
MW was supported by the California Institute of Technology and 
\'Ecole normale sup\'erieure Paris.
LS and SP were supported by an NSF REU Site grant. 
We thank the LIGO Scientific Collaboration for permission to use 
data from the time of GRB 031108 for our tests.
LIGO was constructed by the California Institute of Technology and 
Massachusetts Institute of Technology with funding from the National 
Science Foundation and operates under cooperative agreement PHY-0107417. 
This paper has been assigned LIGO Document No.~{LIGO-P09}00097-v4.

\bibliography{P0900097}

\begin{thebibliography}{50}
\expandafter\ifx\csname natexlab\endcsname\relax\def\natexlab#1{#1}\fi
\expandafter\ifx\csname bibnamefont\endcsname\relax
  \def\bibnamefont#1{#1}\fi
\expandafter\ifx\csname bibfnamefont\endcsname\relax
  \def\bibfnamefont#1{#1}\fi
\expandafter\ifx\csname citenamefont\endcsname\relax
  \def\citenamefont#1{#1}\fi
\expandafter\ifx\csname url\endcsname\relax
  \def\url#1{\texttt{#1}}\fi
\expandafter\ifx\csname urlprefix\endcsname\relax\def\urlprefix{URL }\fi
\providecommand{\bibinfo}[2]{#2}
\providecommand{\eprint}[2][]{\url{#2}}

\bibitem[{\citenamefont{{Ott}}(2009)}]{Ott:2008wt}
\bibinfo{author}{\bibfnamefont{C.~D.} \bibnamefont{{Ott}}},
  \bibinfo{journal}{Classical and Quantum Gravity}
  \textbf{\bibinfo{volume}{26}}, \bibinfo{pages}{063001}
  (\bibinfo{year}{2009}), \eprint{0809.0695}.

\bibitem[{\citenamefont{Cutler et~al.}(1993)\citenamefont{Cutler, Apostolatos,
  Bildsten, Finn, Flanagan, Kennefick, Markovic, Ori, Poisson, Sussman
  et~al.}}]{PhysRevLett.70.2984}
\bibinfo{author}{\bibfnamefont{C.}~\bibnamefont{Cutler}},
  \bibinfo{author}{\bibfnamefont{T.~A.} \bibnamefont{Apostolatos}},
  \bibinfo{author}{\bibfnamefont{L.}~\bibnamefont{Bildsten}},
  \bibinfo{author}{\bibfnamefont{L.~S.} \bibnamefont{Finn}},
  \bibinfo{author}{\bibfnamefont{E.~E.} \bibnamefont{Flanagan}},
  \bibinfo{author}{\bibfnamefont{D.}~\bibnamefont{Kennefick}},
  \bibinfo{author}{\bibfnamefont{D.~M.} \bibnamefont{Markovic}},
  \bibinfo{author}{\bibfnamefont{A.}~\bibnamefont{Ori}},
  \bibinfo{author}{\bibfnamefont{E.}~\bibnamefont{Poisson}},
  \bibinfo{author}{\bibfnamefont{G.~J.} \bibnamefont{Sussman}},
  \bibnamefont{et~al.}, \bibinfo{journal}{Phys. Rev. Lett.}
  \textbf{\bibinfo{volume}{70}}, \bibinfo{pages}{2984} (\bibinfo{year}{1993}).

\bibitem[{\citenamefont{M{\'{e}}sz{\'{a}}ros}(2006)}]{Meszaros:2006rc}
\bibinfo{author}{\bibfnamefont{P.}~\bibnamefont{M{\'{e}}sz{\'{a}}ros}},
  \bibinfo{journal}{Rept. Prog. Phys.} \textbf{\bibinfo{volume}{69}},
  \bibinfo{pages}{2259} (\bibinfo{year}{2006}), \eprint{astro-ph/0605208}.

\bibitem[{\citenamefont{Cutler and Thorne}(2002)}]{CuTh:02}
\bibinfo{author}{\bibfnamefont{C.}~\bibnamefont{Cutler}} \bibnamefont{and}
  \bibinfo{author}{\bibfnamefont{K.~S.} \bibnamefont{Thorne}}
  (\bibinfo{year}{2002}), \eprint{gr-qc/0204090}.

\bibitem[{\citenamefont{Bloom et~al.}(2009)}]{Bloom:2009vx}
\bibinfo{author}{\bibfnamefont{J.~S.} \bibnamefont{Bloom}} \bibnamefont{et~al.}
  (\bibinfo{year}{2009}), \eprint{0902.1527}.

\bibitem[{\citenamefont{Kanner et~al.}(2008)}]{Kanner:2008zh}
\bibinfo{author}{\bibfnamefont{J.}~\bibnamefont{Kanner}} \bibnamefont{et~al.},
  \bibinfo{journal}{Class. Quant. Grav.} \textbf{\bibinfo{volume}{25}},
  \bibinfo{pages}{184034} (\bibinfo{year}{2008}), \eprint{0803.0312}.

\bibitem[{\citenamefont{Van~Elewyck et~al.}(2009)}]{VanElewyck:2009pf}
\bibinfo{author}{\bibfnamefont{V.}~\bibnamefont{Van~Elewyck}}
  \bibnamefont{et~al.}, \bibinfo{journal}{Int. J. Mod. Phys.}
  \textbf{\bibinfo{volume}{D18}}, \bibinfo{pages}{1655} (\bibinfo{year}{2009}),
  \eprint{0906.4957}.

\bibitem[{\citenamefont{Abbott et~al.}(2009{\natexlab{a}})}]{S5firstyear}
\bibinfo{author}{\bibfnamefont{B.~P.} \bibnamefont{Abbott}}
  \bibnamefont{et~al.}, \bibinfo{journal}{Phys. Rev. D}
  \textbf{\bibinfo{volume}{80}}, \bibinfo{pages}{102001}
  (\bibinfo{year}{2009}{\natexlab{a}}), \eprint{0905.0020}.

\bibitem[{\citenamefont{Abbott et~al.}(2009{\natexlab{b}})}]{Abbott:2009tt}
\bibinfo{author}{\bibfnamefont{B.~P.} \bibnamefont{Abbott}}
  \bibnamefont{et~al.}, \bibinfo{journal}{Phys. Rev.}
  \textbf{\bibinfo{volume}{D79}}, \bibinfo{pages}{122001}
  (\bibinfo{year}{2009}{\natexlab{b}}), \eprint{0901.0302}.

\bibitem[{\citenamefont{Abbott et~al.}(2008{\natexlab{a}})}]{Abbott:2007rh}
\bibinfo{author}{\bibfnamefont{B.}~\bibnamefont{Abbott}} \bibnamefont{et~al.},
  \bibinfo{journal}{\apj} \textbf{\bibinfo{volume}{681}}, \bibinfo{pages}{1419}
  (\bibinfo{year}{2008}{\natexlab{a}}), \eprint{0711.1163}.

\bibitem[{\citenamefont{Blackburn et~al.}(2008)}]{Blackburn:2008ah}
\bibinfo{author}{\bibfnamefont{L.}~\bibnamefont{Blackburn}}
  \bibnamefont{et~al.}, \bibinfo{journal}{Class. Quant. Grav.}
  \textbf{\bibinfo{volume}{25}}, \bibinfo{pages}{184004}
  (\bibinfo{year}{2008}), \eprint{0804.0800}.

\bibitem[{\citenamefont{Leroy}(2009)}]{Leroy:2009zz}
\bibinfo{author}{\bibfnamefont{N.}~\bibnamefont{Leroy}}
  (\bibinfo{collaboration}{LIGO Scientific Collaboration and Virgo
  Collaboration}), \bibinfo{journal}{Class. Quant. Grav.}
  \textbf{\bibinfo{volume}{26}}, \bibinfo{pages}{204007}
  (\bibinfo{year}{2009}).

\bibitem[{Xpi()}]{Xpipeline}
\bibinfo{note}{\url{https://geco.phys.columbia.edu/xpipeline/wiki}}.

\bibitem[{\citenamefont{Chatterji et~al.}(2006)\citenamefont{Chatterji,
  Lazzarini, Stein, Sutton, Searle, and Tinto}}]{Ch_etal:06}
\bibinfo{author}{\bibfnamefont{S.}~\bibnamefont{Chatterji}},
  \bibinfo{author}{\bibfnamefont{A.}~\bibnamefont{Lazzarini}},
  \bibinfo{author}{\bibfnamefont{L.}~\bibnamefont{Stein}},
  \bibinfo{author}{\bibfnamefont{P.}~\bibnamefont{Sutton}},
  \bibinfo{author}{\bibfnamefont{A.}~\bibnamefont{Searle}}, \bibnamefont{and}
  \bibinfo{author}{\bibfnamefont{M.}~\bibnamefont{Tinto}},
  \bibinfo{journal}{Phys. Rev. D} \textbf{\bibinfo{volume}{74}},
  \bibinfo{pages}{082005} (\bibinfo{year}{2006}).

\bibitem[{\citenamefont{Abbott et~al.}(2010)}]{P0900023}
\bibinfo{author}{\bibfnamefont{B.~P.} \bibnamefont{Abbott}}
  \bibnamefont{et~al.}, \bibinfo{journal}{The Astrophysical Journal, to appear}
   (\bibinfo{year}{2010}), \eprint{0908.3824}.

\bibitem[{\citenamefont{Cannon}(2008)}]{0264-9381-25-10-105024}
\bibinfo{author}{\bibfnamefont{K.~C.} \bibnamefont{Cannon}},
  \bibinfo{journal}{Classical and Quantum Gravity}
  \textbf{\bibinfo{volume}{25}}, \bibinfo{pages}{105024 (13pp)}
  (\bibinfo{year}{2008}).

\bibitem[{\citenamefont{Anderson et~al.}(2001)\citenamefont{Anderson, Brady,
  Creighton, and Flanagan}}]{AnBrCrFl:01}
\bibinfo{author}{\bibfnamefont{W.~G.} \bibnamefont{Anderson}},
  \bibinfo{author}{\bibfnamefont{P.~R.} \bibnamefont{Brady}},
  \bibinfo{author}{\bibfnamefont{J.~D.~E.} \bibnamefont{Creighton}},
  \bibnamefont{and} \bibinfo{author}{\bibfnamefont{E.~E.}
  \bibnamefont{Flanagan}}, \bibinfo{journal}{Phys. Rev. D}
  \textbf{\bibinfo{volume}{63}}, \bibinfo{pages}{042003}
  (\bibinfo{year}{2001}).

\bibitem[{\citenamefont{Sylvestre}(2002)}]{Sy:02}
\bibinfo{author}{\bibfnamefont{J.}~\bibnamefont{Sylvestre}},
  \bibinfo{journal}{Phys. Rev. D} \textbf{\bibinfo{volume}{66}},
  \bibinfo{pages}{102004} (\bibinfo{year}{2002}).

\bibitem[{\citenamefont{Gursel and Tinto}(1989)}]{GuTi:89}
\bibinfo{author}{\bibfnamefont{Y.}~\bibnamefont{Gursel}} \bibnamefont{and}
  \bibinfo{author}{\bibfnamefont{M.}~\bibnamefont{Tinto}},
  \bibinfo{journal}{Phys. Rev. D} \textbf{\bibinfo{volume}{40}},
  \bibinfo{pages}{3884} (\bibinfo{year}{1989}).

\bibitem[{\citenamefont{Flanagan and Hughes}(1998)}]{FlHu:98b}
\bibinfo{author}{\bibfnamefont{E.~E.} \bibnamefont{Flanagan}} \bibnamefont{and}
  \bibinfo{author}{\bibfnamefont{S.~A.} \bibnamefont{Hughes}},
  \bibinfo{journal}{Phys. Rev. D} \textbf{\bibinfo{volume}{57}},
  \bibinfo{pages}{4566} (\bibinfo{year}{1998}).

\bibitem[{\citenamefont{Klimenko et~al.}(2005)\citenamefont{Klimenko, Mohanty,
  Rakhmanov, and Mitselmakher}}]{KlMoRaMi:05}
\bibinfo{author}{\bibfnamefont{S.}~\bibnamefont{Klimenko}},
  \bibinfo{author}{\bibfnamefont{S.}~\bibnamefont{Mohanty}},
  \bibinfo{author}{\bibfnamefont{M.}~\bibnamefont{Rakhmanov}},
  \bibnamefont{and}
  \bibinfo{author}{\bibfnamefont{G.}~\bibnamefont{Mitselmakher}},
  \bibinfo{journal}{Phys. Rev. D} \textbf{\bibinfo{volume}{72}},
  \bibinfo{pages}{122002} (\bibinfo{year}{2005}).

\bibitem[{\citenamefont{Klimenko et~al.}(2006)\citenamefont{Klimenko, Mohanty,
  Rakhmanov, and Mitselmakher}}]{KlMoRaMi:06}
\bibinfo{author}{\bibfnamefont{S.}~\bibnamefont{Klimenko}},
  \bibinfo{author}{\bibfnamefont{S.}~\bibnamefont{Mohanty}},
  \bibinfo{author}{\bibfnamefont{M.}~\bibnamefont{Rakhmanov}},
  \bibnamefont{and}
  \bibinfo{author}{\bibfnamefont{G.}~\bibnamefont{Mitselmakher}},
  \bibinfo{journal}{J. Phys. Conf. Ser.} \textbf{\bibinfo{volume}{32}},
  \bibinfo{pages}{12} (\bibinfo{year}{2006}).

\bibitem[{\citenamefont{Mohanty et~al.}(2006)\citenamefont{Mohanty, Rakhmanov,
  Klimenko, and Mitselmakher}}]{MoRaKlMi:06}
\bibinfo{author}{\bibfnamefont{S.}~\bibnamefont{Mohanty}},
  \bibinfo{author}{\bibfnamefont{M.}~\bibnamefont{Rakhmanov}},
  \bibinfo{author}{\bibfnamefont{S.}~\bibnamefont{Klimenko}}, \bibnamefont{and}
  \bibinfo{author}{\bibfnamefont{G.}~\bibnamefont{Mitselmakher}},
  \bibinfo{journal}{Class. Quant. Grav.} \textbf{\bibinfo{volume}{23}},
  \bibinfo{pages}{4799} (\bibinfo{year}{2006}), \eprint{gr-qc/0601076}.

\bibitem[{\citenamefont{Rakhmanov}(2006)}]{Ra:06}
\bibinfo{author}{\bibfnamefont{M.}~\bibnamefont{Rakhmanov}},
  \bibinfo{journal}{Class. Quant. Grav.} \textbf{\bibinfo{volume}{23}},
  \bibinfo{pages}{S673} (\bibinfo{year}{2006}), \eprint{gr-qc/0604005}.

\bibitem[{\citenamefont{Klimenko et~al.}(2008)\citenamefont{Klimenko, Yakushin,
  Mercer, and Mitselmakher}}]{Klimenko:2008fu}
\bibinfo{author}{\bibfnamefont{S.}~\bibnamefont{Klimenko}},
  \bibinfo{author}{\bibfnamefont{I.}~\bibnamefont{Yakushin}},
  \bibinfo{author}{\bibfnamefont{A.}~\bibnamefont{Mercer}}, \bibnamefont{and}
  \bibinfo{author}{\bibfnamefont{G.}~\bibnamefont{Mitselmakher}},
  \bibinfo{journal}{Class. Quant. Grav.} \textbf{\bibinfo{volume}{25}},
  \bibinfo{pages}{114029} (\bibinfo{year}{2008}), \eprint{0802.3232}.

\bibitem[{\citenamefont{Searle et~al.}(2008)\citenamefont{Searle, Sutton,
  Tinto, and Woan}}]{Searle:2007uv}
\bibinfo{author}{\bibfnamefont{A.~C.} \bibnamefont{Searle}},
  \bibinfo{author}{\bibfnamefont{P.~J.} \bibnamefont{Sutton}},
  \bibinfo{author}{\bibfnamefont{M.}~\bibnamefont{Tinto}}, \bibnamefont{and}
  \bibinfo{author}{\bibfnamefont{G.}~\bibnamefont{Woan}},
  \bibinfo{journal}{Class. Quant. Grav.} \textbf{\bibinfo{volume}{25}},
  \bibinfo{pages}{114038} (\bibinfo{year}{2008}), \eprint{0712.0196}.

\bibitem[{\citenamefont{Searle et~al.}(2009)\citenamefont{Searle, Sutton, and
  Tinto}}]{Searle:2008ap}
\bibinfo{author}{\bibfnamefont{A.~C.} \bibnamefont{Searle}},
  \bibinfo{author}{\bibfnamefont{P.~J.} \bibnamefont{Sutton}},
  \bibnamefont{and} \bibinfo{author}{\bibfnamefont{M.}~\bibnamefont{Tinto}},
  \bibinfo{journal}{Class. Quant. Grav.} \textbf{\bibinfo{volume}{26}},
  \bibinfo{pages}{155017} (\bibinfo{year}{2009}), \eprint{0809.2809}.

\bibitem[{\citenamefont{Wen and Schutz}(2005)}]{WeSc:05}
\bibinfo{author}{\bibfnamefont{L.}~\bibnamefont{Wen}} \bibnamefont{and}
  \bibinfo{author}{\bibfnamefont{B.}~\bibnamefont{Schutz}},
  \bibinfo{journal}{Class. Quantum Grav.} \textbf{\bibinfo{volume}{22}},
  \bibinfo{pages}{S1321} (\bibinfo{year}{2005}).

\bibitem[{\citenamefont{Hayama et~al.}(2007)\citenamefont{Hayama, Mohanty,
  Rakhmanov, and Desai}}]{Hayama:2007iz}
\bibinfo{author}{\bibfnamefont{K.}~\bibnamefont{Hayama}},
  \bibinfo{author}{\bibfnamefont{S.~D.} \bibnamefont{Mohanty}},
  \bibinfo{author}{\bibfnamefont{M.}~\bibnamefont{Rakhmanov}},
  \bibnamefont{and} \bibinfo{author}{\bibfnamefont{S.}~\bibnamefont{Desai}},
  \bibinfo{journal}{Class. Quant. Grav.} \textbf{\bibinfo{volume}{24}},
  \bibinfo{pages}{S681} (\bibinfo{year}{2007}), \eprint{0709.0940}.

\bibitem[{\citenamefont{Tinto}(1996)}]{Tinto:96}
\bibinfo{author}{\bibfnamefont{M.}~\bibnamefont{Tinto}}, in
  \emph{\bibinfo{booktitle}{Proceedings of the International Conference on
  Gravitational Waves: Source and Detectors.}} (\bibinfo{publisher}{World
  Scientific (Singapore)}, \bibinfo{year}{1996}).

\bibitem[{\citenamefont{Sylvestre}(2003)}]{Sy:03}
\bibinfo{author}{\bibfnamefont{J.}~\bibnamefont{Sylvestre}},
  \bibinfo{journal}{Phys. Rev. D} \textbf{\bibinfo{volume}{68}},
  \bibinfo{pages}{102005} (\bibinfo{year}{2003}).

\bibitem[{\citenamefont{Summerscales et~al.}(2008)\citenamefont{Summerscales,
  Burrows, Finn, and Ott}}]{Summerscales:2007xq}
\bibinfo{author}{\bibfnamefont{T.~Z.} \bibnamefont{Summerscales}},
  \bibinfo{author}{\bibfnamefont{A.}~\bibnamefont{Burrows}},
  \bibinfo{author}{\bibfnamefont{L.~S.} \bibnamefont{Finn}}, \bibnamefont{and}
  \bibinfo{author}{\bibfnamefont{C.~D.} \bibnamefont{Ott}},
  \bibinfo{journal}{The Astrophysical Journal} \textbf{\bibinfo{volume}{678}},
  \bibinfo{pages}{1142} (\bibinfo{year}{2008}), \eprint{0704.2157}.

\bibitem[{\citenamefont{Cadonati}(2004)}]{Ca:04}
\bibinfo{author}{\bibfnamefont{L.}~\bibnamefont{Cadonati}},
  \bibinfo{journal}{Class. Quantum Grav.} \textbf{\bibinfo{volume}{21}},
  \bibinfo{pages}{S1695} (\bibinfo{year}{2004}).

\bibitem[{\citenamefont{Fairhurst}(2009)}]{Fa:09}
\bibinfo{author}{\bibfnamefont{S.}~\bibnamefont{Fairhurst}},
  \bibinfo{journal}{New Journal of Physics} \textbf{\bibinfo{volume}{11}},
  \bibinfo{pages}{123006} (\bibinfo{year}{2009}).

\bibitem[{\citenamefont{Flanagan and Hughes}(2005)}]{Flanagan:2005yc}
\bibinfo{author}{\bibfnamefont{E.~E.} \bibnamefont{Flanagan}} \bibnamefont{and}
  \bibinfo{author}{\bibfnamefont{S.~A.} \bibnamefont{Hughes}},
  \bibinfo{journal}{{New Journal of Physics}} \textbf{\bibinfo{volume}{7}},
  \bibinfo{pages}{204} (\bibinfo{year}{2005}), \eprint{gr-qc/0501041}.

\bibitem[{\citenamefont{Abbott et~al.}(2004)}]{abbottnim04}
\bibinfo{author}{\bibfnamefont{B.}~\bibnamefont{Abbott}} \bibnamefont{et~al.},
  \bibinfo{journal}{Nucl. Inst. \& Meth. in Phys. Res.}
  \textbf{\bibinfo{volume}{517}}, \bibinfo{pages}{154} (\bibinfo{year}{2004}).

\bibitem[{\citenamefont{Abbott et~al.}(2009{\natexlab{c}})}]{abbott-2007}
\bibinfo{author}{\bibfnamefont{B.}~\bibnamefont{Abbott}} \bibnamefont{et~al.},
  \bibinfo{journal}{Rep. Prog. Phys.} \textbf{\bibinfo{volume}{72}},
  \bibinfo{pages}{076901} (\bibinfo{year}{2009}{\natexlab{c}}),
  \eprint{arXiv:0711.3041}, \urlprefix\url{doi:10.1088/0034-4885/72/7/076901}.

\bibitem[{\citenamefont{Acernese et~al.}(2008)}]{virgo08}
\bibinfo{author}{\bibfnamefont{F.}~\bibnamefont{Acernese}}
  \bibnamefont{et~al.}, \bibinfo{journal}{Classical and Quantum Gravity}
  \textbf{\bibinfo{volume}{25}}, \bibinfo{pages}{114045}
  (\bibinfo{year}{2008}).

\bibitem[{\citenamefont{Grote et~al.}(2008)}]{geo08}
\bibinfo{author}{\bibfnamefont{H.}~\bibnamefont{Grote}} \bibnamefont{et~al.},
  \bibinfo{journal}{Classical and Quantum Gravity}
  \textbf{\bibinfo{volume}{25}}, \bibinfo{pages}{114043}
  (\bibinfo{year}{2008}).

\bibitem[{\citenamefont{Chatterji et~al.}(2004)\citenamefont{Chatterji,
  Blackburn, Martin, and Katsavounidis}}]{ChBlMaKa:04}
\bibinfo{author}{\bibfnamefont{S.}~\bibnamefont{Chatterji}},
  \bibinfo{author}{\bibfnamefont{L.}~\bibnamefont{Blackburn}},
  \bibinfo{author}{\bibfnamefont{G.}~\bibnamefont{Martin}}, \bibnamefont{and}
  \bibinfo{author}{\bibfnamefont{E.}~\bibnamefont{Katsavounidis}},
  \bibinfo{journal}{Class. Quant. Grav.} \textbf{\bibinfo{volume}{21}},
  \bibinfo{pages}{S1809} (\bibinfo{year}{2004}).

\bibitem[{\citenamefont{Abbott et~al.}(2005)}]{Ab_etal:05c}
\bibinfo{author}{\bibfnamefont{B.}~\bibnamefont{Abbott}} \bibnamefont{et~al.},
  \bibinfo{journal}{Phys. Rev. D} \textbf{\bibinfo{volume}{72}},
  \bibinfo{pages}{042002} (\bibinfo{year}{2005}), \eprint{gr-qc/0501068}.

\bibitem[{\citenamefont{Abbott et~al.}(2008{\natexlab{b}})}]{Ab_etal:08}
\bibinfo{author}{\bibfnamefont{B.}~\bibnamefont{Abbott}} \bibnamefont{et~al.},
  \bibinfo{journal}{Phys. Rev. D} \textbf{\bibinfo{volume}{77}},
  \bibinfo{pages}{062004} (\bibinfo{year}{2008}{\natexlab{b}}),
  \eprint{gr-qc/0709.0766}.

\bibitem[{\citenamefont{Brady et~al.}(2004)\citenamefont{Brady, Creighton, and
  Wiseman}}]{Brady:2004gt}
\bibinfo{author}{\bibfnamefont{P.~R.} \bibnamefont{Brady}},
  \bibinfo{author}{\bibfnamefont{J.~D.~E.} \bibnamefont{Creighton}},
  \bibnamefont{and} \bibinfo{author}{\bibfnamefont{A.~G.}
  \bibnamefont{Wiseman}}, \bibinfo{journal}{Class. Quant. Grav.}
  \textbf{\bibinfo{volume}{21}}, \bibinfo{pages}{S1775} (\bibinfo{year}{2004}),
  \eprint{gr-qc/0405044}.

\bibitem[{\citenamefont{Biswas et~al.}(2009)\citenamefont{Biswas, Brady,
  Creighton, and Fairhurst}}]{Biswas:2007ni}
\bibinfo{author}{\bibfnamefont{R.}~\bibnamefont{Biswas}},
  \bibinfo{author}{\bibfnamefont{P.~R.} \bibnamefont{Brady}},
  \bibinfo{author}{\bibfnamefont{J.~D.~E.} \bibnamefont{Creighton}},
  \bibnamefont{and}
  \bibinfo{author}{\bibfnamefont{S.}~\bibnamefont{Fairhurst}},
  \bibinfo{journal}{Classical and Quantum Gravity}
  \textbf{\bibinfo{volume}{26}}, \bibinfo{pages}{175009 (19pp)}
  (\bibinfo{year}{2009}), \eprint{0710.0465}.

\bibitem[{\citenamefont{Abbott
  et~al.}(2008{\natexlab{c}})}]{PhysRevLett.101.211102}
\bibinfo{author}{\bibfnamefont{B.}~\bibnamefont{Abbott}} \bibnamefont{et~al.},
  \bibinfo{journal}{Phys. Rev. Lett.} \textbf{\bibinfo{volume}{101}},
  \bibinfo{pages}{211102} (\bibinfo{year}{2008}{\natexlab{c}}).

\bibitem[{\citenamefont{Riles}(2004)}]{Ri:04}
\bibinfo{author}{\bibfnamefont{K.}~\bibnamefont{Riles}}, \bibinfo{journal}{LIGO
  Document: LIGO-T040055-00-Z}  (\bibinfo{year}{2004}).

\bibitem[{\citenamefont{Hurley et~al.}()}]{GRB031108}
\bibinfo{author}{\bibfnamefont{K.}~\bibnamefont{Hurley}} \bibnamefont{et~al.},
  \bibinfo{note}{\url{http://gcn.gsfc.nasa.gov/gcn3/2441.gcn3}}.

\bibitem[{gcn()}]{gcn}
\bibinfo{note}{\url{http://gcn.gsfc.nasa.gov/}}.

\bibitem[{\citenamefont{{Gehrels} et~al.}(2004)\citenamefont{{Gehrels},
  {Chincarini}, {Giommi}, {Mason}, {Nousek}, {Wells}, {White}, {Barthelmy},
  {Burrows}, {Cominsky} et~al.}}]{swift04}
\bibinfo{author}{\bibfnamefont{N.}~\bibnamefont{{Gehrels}}},
  \bibinfo{author}{\bibfnamefont{G.}~\bibnamefont{{Chincarini}}},
  \bibinfo{author}{\bibfnamefont{P.}~\bibnamefont{{Giommi}}},
  \bibinfo{author}{\bibfnamefont{K.~O.} \bibnamefont{{Mason}}},
  \bibinfo{author}{\bibfnamefont{J.~A.} \bibnamefont{{Nousek}}},
  \bibinfo{author}{\bibfnamefont{A.~A.} \bibnamefont{{Wells}}},
  \bibinfo{author}{\bibfnamefont{N.~E.} \bibnamefont{{White}}},
  \bibinfo{author}{\bibfnamefont{S.~D.} \bibnamefont{{Barthelmy}}},
  \bibinfo{author}{\bibfnamefont{D.~N.} \bibnamefont{{Burrows}}},
  \bibinfo{author}{\bibfnamefont{L.~R.} \bibnamefont{{Cominsky}}},
  \bibnamefont{et~al.}, \bibinfo{journal}{\apj} \textbf{\bibinfo{volume}{611}},
  \bibinfo{pages}{1005} (\bibinfo{year}{2004}).

\bibitem[{fer()}]{fermi}
\bibinfo{note}{\url{http://fermi.gsfc.nasa.gov/}}.

\end{thebibliography}
\end{document}